\documentstyle{ar_prepr}

\input epsf.tex    

\input psfig.sty

\def\stacksymbols #1#2#3#4{\def\theguybelow{#2}
        \def\verticalposition{\lower#3pt}
        \def\spacingwithinsymbol{\baselineskip0pt\lineskip#4pt}
        \mathrel{\mathpalette\intermediary#1}}
\def\intermediary #1#2{\verticalposition\vbox{\spacingwithinsymbol
        \everycr={}\tabskip0pt
        \halign{$\mathsurround0pt#1\hfil##\hfil$\crcr#2\crcr
                \theguybelow\crcr}}}
\def\lta{\stacksymbols{<}{\sim}{2.5}{.2}}
\def\gta{\stacksymbols{>}{\sim}{3}{.5}}

\newcommand\apj{{\it Ap. J. }}
\newcommand\apjs{{\it Ap. J. Suppl. }}
\newcommand\aj{{\it Astron. J. }}
\newcommand\araa{{\it Annu. Rev. Astron. Astrophys. }}
\newcommand\mnras{{\it MNRAS }}
\newcommand\aea{{\it Astron. Astrophys. }}
\newcommand\aeas{{\it Astron. Astrophys. Suppl.}}

\begin{document}

%

\jname{..}
\jyear{}
\jvol{}
\ARinfo{}

\title{Hot Gas In and Around Elliptical Galaxies}

\markboth{MATHEWS \& BRIGHENTI}{HOT GAS IN AND AROUND ELLIPTICAL GALAXIES}

\author{William G. Mathews$^1$ and Fabrizio Brighenti$^{1,2}$
\affiliation{$^1$University of California Observatories/Lick
Observatory,
Department of Astronomy and Astrophysics,
University of California, Santa Cruz, CA 95064
mathews@ucolick.org
$^2$Dipartimento di Astronomia,
Universit\`a di Bologna,
via Ranzani 1,
Bologna 40127, Italy
brighenti@bo.astro.it}}


\begin{abstract}
We review the origin, evolution and physical nature of
hot gas in elliptical galaxies and associated
galaxy groups.
Unanticipated recent X-ray observations with
Chandra and XMM indicate much less cooling than
previously expected. 
Consequently, many long-held assumptions
need to be reexamined or discarded and new approaches 
must be explored.
Chief among these are the role of heating by active
galactic nuclei, 
the influence of radio lobes on the hot gas, 
details of the cooling process, 
possible relation between the hot and colder gas 
in elliptical galaxies, 
and the complexities of stellar enrichment of the hot gas.
\end{abstract}

\maketitle

\section{INTRODUCTION}

Recent Chandra and XMM X-ray observations of the hot gas in
elliptical galaxies and galaxy clusters
have radically upset the traditional
concept of cooling flows.
On the other hand, these same observations
may reveal new ways of understanding the origin
and evolution of this gas.
The most unexpected new discovery
has been the absence in XMM spectra of
emission from gas at intermediate or low temperatures,
implying that the cooling gas is somehow hidden from view or that
the cooling rate is much less than previously thought.
Chandra X-ray images show that the
hot gas deep inside E and cD galaxies near
the centers of cooling flows is often highly disturbed.
Evidently, massive black holes, thought to
inhabit the cores of all stellar bulges,
may become energy sources
when stimulated by inflowing hot gas.

Can the central heating visible in Chandra images
explain the absence of cooling in XMM spectra?
This is one of many fundamental cooling flow questions
that we review here.
There is also a problem with terminology.
Since the spectral evidence for cooling has weakened,
the ``cooling flow'' concept may no longer be appropriate.
Nevertheless, in keeping with accepted usage
we occasionally use ``cooling flow'' when ``galactic flow''
might be preferred.
In any case,
the hot gas in and around elliptical galaxies
is of particular astronomical interest
because its origin and metal enrichment are closely
entwined with the merger origin of central elliptical (E) galaxies
in galaxy groups.
A better understanding of the gas is likely to
clarify baryonic structure formation in general.
Fortunately, even with our currently
incomplete understanding of its origin and evolution,
the hot gas can contribute important information about
the stellar kinematics and mass to light ratios in elliptical
galaxies.

This review is necessarily limited to elliptical galaxies
of moderate or high luminosity.
X-ray emission from hot gas in low luminosity E galaxies
is difficult to observe since it is
masked by X-rays from stellar sources.
We also emphasize whenever possible those X-ray ellipticals that
are relatively undisturbed by recent mergers, stripping interactions
with hot cluster gas or powerful radio sources.
Although the X-ray properties of S0 galaxies 
can resemble those of E galaxies,
we restrict our discussion to E galaxies owing to 
the uniformity of their optical structures and their less
controversial origin.

The evolution and physical condition of
the hot gas in rich clusters of galaxies
and in individual E galaxies are similar in many ways.
But there are some important distinctions.
Because smaller structures are often older in our hierarchical
universe,
the hot gas in elliptical galaxies and their associated galaxy
groups may on average
be less disturbed by ongoing mergers than their
younger, more massive cluster counterparts.
X-ray bright E galaxies that are rather isolated and, one hopes,
undisturbed may contain in their metallicity gradient
unique information about star and galaxy formation in
the distant past.
This vital information is lost when such
groups merge and mix to form rich clusters.
Another key attribute of hot gas on galactic scales
is the possible importance of stellar mass loss and continued
Type Ia supernova activity in supplying and enriching the hot gas.
Finally, the properties of the hot gas can be determined with
more confidence on galactic scales where the gravitational
potential is better known from the stellar light.

Notwithstanding their astrophysical importance,
proximity and relative ease
of interpretation, far less
Chandra and XMM data are available for E galaxies
and their groups than for galaxy clusters. 
Both telescopes were launched in 1999.
As we write this review in the Fall of 2002 
no XMM spectrum of a bright, relatively isolated,
group-centered elliptical
galaxy has appeared in the standard journals, 
but we expect this data to appear very soon.

Our review is not intended to be a comprehensive review of
all the literature that has appeared since the excellent previous
cooling flow reviews by
Sarazin (1986),
Fabbiano (1989),
Sarazin (1990) and
Fabian (1994).
This is due in part to the rapid increase in the quality of
the observations which tend to eclipse previous studies
only a few years older.
Our primary emphasis is to compare observations
with theoretical expectations and to bring attention to
the frequent dissonance between the two.
This approach is complementary to the recent ARAA
review of the X-ray properties of galaxy groups by
Mulchaey (2000) who presents a more complete
discussion of the observations.

\section{OVERVIEW OF HOT GAS FLOWS IN AND NEAR ELLIPTICAL GALAXIES}

Optical evidence indicates that many or most
massive elliptical galaxies
formed at high redshifts ($z \gta 2$)
and there is strong theoretical evidence that they formed
by mergers in group environments.
Later, many ellipticals and their groups
merged into massive galaxy clusters.
Those that did not,
the so-called isolated ellipticals, are surrounded
by unmerged group companions and hot gas. 
As a consequence of their formation and 
galactic dynamics, big elliptical galaxies are often
found near the centers of groups and rich clusters.
It is natural that the diffuse thermal gas contained in these hot
stellar systems is also hot. 

There are two main sources of hot gas in elliptical
galaxies: internal and external.
Evolving stars inside the elliptical galaxy
continuously eject gas at a rate
$\sim 1.3[L_B/(10^{11} L_{B,\odot})]$ $M_{\odot}$ yr$^{-1}$.
It is generally assumed that
gas ejected by orbiting red giant stars 
passes through shocks and is raised to the
stellar kinematic temperature $T_* \approx T_{vir} \approx
\mu m_p \sigma^2 /k \sim 10^7~{\rm K} \sim 1$ keV
where $\mu$ is the molecular weight 
and $\sigma$ is the stellar velocity dispersion. 
Type Ia supernovae provide some additional heating. 
The large X-ray luminosities of massive E galaxies,
$L_x \sim 10^{40} - 10^{43}$ ergs s$^{-1}$ 
for $L_B > 3 \times 10^{10}$ $L_{B,\odot}$,
indicate that most of the internally produced gas is 
currently trapped in the galactic or group potential. 
But at early times, when most of the galactic stars were forming,
Type II supernovae were frequent enough to drive winds
of enriched gas into the local environment.
Gas expelled in this manner from both
central and non-central group or cluster
galaxies has enriched the hot gas far beyond the stellar image
of the central luminous E or cD galaxy.
In time, some of 
this local (circumgalactic) gas flows back into the
central galaxy, providing an external source of gas.
Continued accretion from the
ambient cosmological flow that is
gravitationally bound to the group or cluster
is an additional source of external gas.
As diffuse external gas arrives
after having fallen through the deeper potential well
of the surrounding group/cluster, it is shock-heated to
the virial temperature of the galaxy group/cluster.
This more distant accreted and shocked gas is hotter
than gas virialized to $T_*$ 
deeper in the stellar potential
of the E galaxy, and the two together form an 
an outwardly 
increasing gas temperature that is commonly observed.
The accumulated mass of circumgalactic gas with $T > T_*$,
bound to the dark matter halo, can extend far beyond
the optical image of the luminous E galaxy.

The electron density of 
the hot gas in giant elliptical galaxies 
is typically $n(0) \sim 0.1$ cm$^{-3}$ 
at the center and  
declines with radius as $n \propto r^{-1.25 \pm 0.25}$.
Depending on its spatial extent, the total mass of hot 
gas in massive E galaxies varies up to at least 
several $10^{10}$ $M_{\odot}$
or about $\lta 1$ percent of the total stellar mass. 
This is only a few times less than 
the gas to stellar mass ratio in the Milky Way.
The iron abundance in the hot gas in ellipticals 
increases from $z_{Fe} \sim 0.2 - 0.4$  
beyond the optical image to $z_{Fe} \sim 1 - 2$ in the center
where it is evidently enriched by Type Ia supernovae.

To a good approximation, the hot gas in and near elliptical
galaxies is in hydrostatic equilibrium. 
Supersonic winds are not common in well-observed 
massive ellipticals 
($L_B \gta 3 \times 10^{10}$ $L_{B,\odot}$) because 
they would have low gas densities 
and much lower X-ray luminosities.
A characteristic feature of the hot gas is that the
the dynamical and sound crossing times are nearly
equal, as expected in hydrostatic equilibrium, 
and both are much less than the radiative cooling time.
Any cooling-induced flow is therefore
highly subsonic, essentially in hydrostatic equilibrium.
This equilibrium can be disturbed by mergers or by
energy released 
in an active galactic nucleus (AGN) 
associated with a supermassive black 
hole in the core of the central elliptical.
Nevertheless,
by assuming hydrostatic equilibrium, the total mass distribution
$M_{tot}(r)$ has been determined for many galaxies and clusters
from X-ray observations.
If a galaxy group has been relatively undisturbed
for many Gyrs, the metal enrichment in the 
hot gas may retain a memory of
the (largely SNII-driven) 
galactic winds that occurred in the distant past.
We are just beginning to exploit this gold mine of information.

In addition to the accretion shock,
hot gas in ellipticals is heated further by
Type Ia supernovae and by dissipation
of mass lost from orbiting stars.
As a consequence of the $\sim 0.7$ $M_{\odot}$ of iron 
that each Type Ia is thought to contribute, 
the observed iron abundance in the hot gas is a measure of the 
supernovae heating for a given flow model.
In low luminosity ellipticals, which have shallower
gravitational potentials, current supernova heating may be 
sufficient to drive a galactic outflow.
But in optically bright, X-ray luminous E galaxies, 
supernova heating is generally unable to balance the
radiative losses from the hot plasma 
that produces the X-ray emission we observe.
Paradoxically, the loss of radiative
energy from the hot gas does not result
in a proportional decrease in the local gas temperature.
Instead, this loss of thermal energy is immediately
compensated by $Pdv$ compression in the
gravitational 
potential of the galaxy/group that maintains the temperature
$\sim T_{vir}$ necessary to support the surrounding
atmosphere of hot gas. 
However, near the center of the flow where there is no deeper
galactic potential, catastrophic radiative cooling
could in principle occur.

But something is wrong, perhaps radically wrong,
with this simple ``cooling flow'' model.
An estimate of the mass cooling rate required to
generate the observed X-ray
luminosity in a bright E galaxy, 
${\dot M} \approx (2 \mu m_p/5 k T)L_{x,bol}$
reveals that several $10^{10}$ $M_{\odot}$ of gas should
have cooled somewhere within
the galaxy over a Hubble time.
The cooling cannot be too concentrated because this mass
exceeds the masses of known central black holes by 
factors of 10 - 20 
and its gravitational attraction on the hot gas 
would produce an unobserved central peak in X-ray emission.
The traditional solution to this problem
is to invoke an {\it ad hoc}
mass ``dropout'' assumption in which the gas somehow cools
throughout a large volume of the flow.
Unfortunately, such distributed cooling cannot result from
thermal instabilities following small perturbations in the
hot gas. 
Instead, larger perturbations formed in 
turbulent regions may be required, but the details remain
uncertain.

Even more astonishing,
detailed X-ray spectra taken with XMM show little
or no emission from ions cooling at temperatures
much below $\sim T_{vir}/2$,
implying that little gas completely cools or perhaps none at all!
There are two possible explanations: (1) the gas is cooling,
but it is not visible and/or (2) the gas is truly not
cooling and the radiative losses are offset by some
source of heating.

In principle, 
X-ray emission from cooling gas can be attenuated or hidden 
(1) by spatially distributed colder gas that absorbs softer 
X-rays or (2) if the cooling is somehow accelerated so that the
X-ray line emission from the cooling gas is reduced.
Although X-ray spectra provide some support for distributed
absorption at energies $\lta 1$ keV, 
the absorbing gas would need to be colder, 
$T_{abs} \lta 10^6$ K, spatially extended and very massive 
$M_{abs} \approx 10^{10} (r/10~{\rm kpc})^2$ $M_{\odot}$.
Emission from this absorbing gas has not been observed.
Alternatively, rapid cooling may be possible 
in localized regions of high metallicity 
such as remnants of Type Ia explosions or 
if cold gas rapidly mixes with the hot gas. 
Cooling may also be very
rapid in the dust-rich gas recently ejected from evolving stars. 
These processes have not been studied in detail. 

The currently most popular
explanation for the absence of spectroscopic
evidence for cooling is that
the hot gas is being heated by active galactic nuclei (AGN) 
in the cores of flow-centered giant ellipticals.
This is reasonable since almost all bright E galaxies have
extended non-thermal radio emission at some level in their cores.
Additional support for this heating hypothesis
is provided by the spectacular
X-ray images from Chandra in which the hot gas
in the central regions of virtually all elliptical galaxies
observed so far 
appears to be highly disturbed and irregular.
In some cases bubbles (X-ray cavities)
filled with relativistic or superheated
gas have displaced the 1 keV thermal gas and appear to
be buoyantly floating upward in the atmosphere.
Many observers have noted with astonishment that the
gas just adjacent to the bubbles is cooler than average,
indicating that the holes
have not been recently produced by strong shocks.

Certainly the luminosities of typical AGN are sufficient
to offset the X-ray luminosities $L_x$ of classic
cooling flows and maintain their gas temperature near $T_{vir}$.
However, the problem with the heating hypothesis,
which is widely unappreciated, is that it is difficult
to communicate this AGN energy to the hot gas at larger radii
and still preserve the globally observed
hot gas temperature and density profiles.

\section{BASIC PROPERTIES OF ELLIPTICAL GALAXIES AND THEIR HOT GAS}

If elliptical galaxies were perfectly homologous stellar systems
with identical stellar populations, then the
``central'' velocity dispersion $\sigma_o$, 
stellar mass $M$ and half-light 
(effective) radius $R_e$ would be related by the virial theorem,
$\sigma_o^2 = \kappa (M/R_e)
= \kappa (L_V/R_e)(M/L_V)$ with constant $\kappa$.
Instead, non-homology and/or stellar population variations conspire 
to place elliptical galaxies on a nearby fundamental plane 
$\sigma_o^2 \propto (L_V/R_e)[R_e^{0.22}\sigma_o^{0.49}]$,
implying that $\kappa (M/L_V) \propto R_e^{0.22} \sigma_o^{0.49}
\propto {M}^{0.24} R_e^{-0.02} \propto {L_V}^{0.32} R_e^{-0.03}$
(Dressler et al. 1987; Djorgovski \& Davis 1987).
The width of the fundamental plane is remarkably small
(Renzini \& Ciotti 1993).
In projection the fundamental plane indicates that 
the binding energy per unit mass decreases with 
stellar (or galactic) mass, $\sigma_o^2 \propto M^{0.32}$
(Faber et al. 1997),
i.e., hot interstellar gas is less bound in low mass ellipticals.
This regularity in the global properties of elliptical galaxies
is useful in interpreting the X-ray 
emission from the hot interstellar gas they contain.

The internal structure is also quite uniform among 
massive E galaxies. 
For those with de Vaucouleurs ($r^{1/4}$) 
stellar profiles the stellar distribution is 
completely determined by the optical half-light radius 
$R_e$ and the total stellar mass $M_{*t}$. 
Provided the dark halos consist of cold, non-interacting 
particles, the halo density profile is determined by 
its virial mass, $M_{dh}$, the mass enclosing a cosmic overdensity 
of $\sim 100$ relative to the critical density 
$\rho_c = 3 H_0^2/8 \pi G$ 
in a flat $\Lambda$CDM universe with $\Omega_m = 0.3$ 
(Eke, Navarro, \& Frenk 1998; Bullock et al. 2000).
If elliptical galaxies formed
by mergers in galaxy groups, as generally believed,
then the hot gas they
contain must be understood in this evolutionary context.
The $r^{1/4}$ stellar
density profile in ellipticals and the high incidence 
($\gta 50$ percent)
of counter-rotating or kinematically independent
stellar cores are natural consequences of
hierarchical mergers (Hernquist \& Barnes 1991; Bender 1996).
While many massive E galaxies currently reside in the
dense cores of rich galaxy clusters, the orbital velocities
of cluster member galaxies are too high for efficient mergers.

In spite of these regularities, 
giant elliptical galaxies come in two flavors 
depending on their mass or optical luminosity. 
Low luminosity ellipticals 
have power law central stellar density profiles, 
disky isophotes, and oblate symmetry consistent 
with their moderate rotation. 
Ellipticals of high luminosity 
have flatter stellar cores within a break radius $r_b$ (typically 
a few percent of $R_e$),
boxy isophotes, and have aspherical 
structures caused by anisotropic stellar velocities, 
not by rotation which is generally small 
(Kormendy \& Bender 1996; Faber et al. 1997; 
Lauer et al. 1998). 
Nevertheless,
many of the most luminous E0 and E1 galaxies are 
thought to be intrinsically spherical to a reasonable approximation 
(Merritt \& Tremblay 1996)
and these galaxies are among the most luminous in X-rays. 
The transition from power-law to core ellipticals
occurs gradually over 
$-20 \gta M_V \gta -22$ where both types coexist
on the same fundamental plane.
Finally, optical spectra of 
ellipticals often show evidence of 
a (sub)population of younger stars, particularly in ellipticals 
with power-law profiles (e.g. Trager et al. 2000; 
Terlevich \& Forbes 2002).

The X-ray luminosities $L_x$ of elliptical galaxies 
correlate with $L_B$ in a manner that 
also exhibits a transition with increasing mass
(Canizares et al. 1987; 
Donnelly et al. 1990; White \& Sarazin 1991; 
Fabbiano, Kim \& Trinchieri 1992;
Eskridge, Fabbiano, \& Kim 1995a,b,c; 
Davis \& White 1996; Brown \& Bregman 1998;
Beuing et al. 1999).
For elliptical galaxies of low luminosity 
(i.e. $L_B \lta L_{B,crit} = 3 \times 10^9$ $L_{B,\odot}$) 
O'Sullivan et al. (2001) find that 
the bolometric X-ray and optical luminosities 
are approximately proportional,  
$L_x \propto L_B$. 
The X-ray emission from these galaxies is apparently dominated by
low-mass X-ray binary stars with 
a different (typically harder) spectrum than that of 
the interstellar gas (Brown \& Bregman 2001). 
The brightest stellar X-ray sources can be 
individually resolved with Chandra 
(e.g. Sarazin, Irwin \& Bregman 2000; 
Blanton, Sarazin, \& Irwin 2001a).
However, the X-ray emission from 
more luminous ellipticals ($L_B \gta L_{B,crit}$) 
varies approximately as $L_x \propto L_B^2$ 
(O'Sullivan et al. 2001), 
clearly indicating a non-stellar origin, 
i.e., the hot gas. 
As shown in Figure 1, 
the scatter about this correlation is enormous, 
$L_x$ varies by almost two orders of magnitude 
for galaxies with similar $L_B$.

\begin{figure}
\centerline{
\psfig{figure=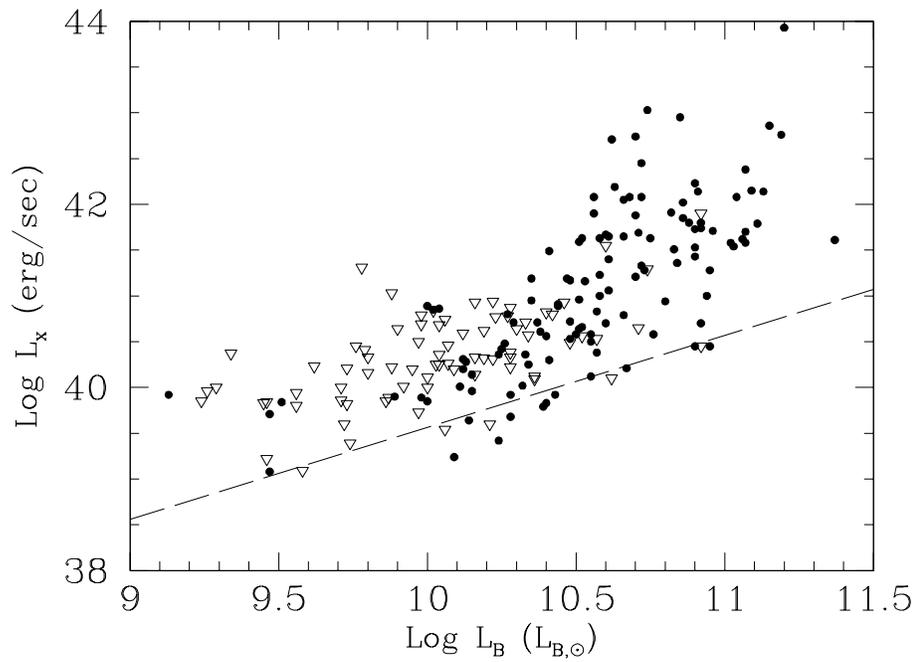,angle=270,width=7.5in} }
\caption{A plot of the bolometric X-ray luminosity
and B-band optical luminosity for elliptical galaxies
(RC3 type T $\leq -4$) from the compilation
of O'Sullivan et al. (2001).
X-ray detections are shown with {\it filled circles} and
upper limits with {\it open triangles}.
The dashed line is an approximate
locus of the total luminosity $L_{x,*} \propto L_B$
of stellar and other discrete sources also from O'Sullivan
et al.}
\label{fig1}
\end{figure}

The large scatter in the $L_x \propto L_B^2$ 
correlation has received much attention but 
an explanation in terms of some specific environmental 
or intrinsic property of the galaxies has been 
elusive 
(e.g. Eskridge, Fabbiano, \& Kim 1995a,b,c).
White and Sarazin (1991) and 
Henriksen \& Cousineau (1999) find that ellipticals 
having other massive galaxies nearby have systematically 
lower $L_x/L_B$ while Brown \& Bregman (2000) find 
a positive correlation with $L_x/L_B$ and the 
local density of galaxies.
Pellegrini (1999) presented evidence that 
ellipticals with power law profiles have much smaller 
range in $L_x/L_B$, as might be expected if 
a larger fraction of 
$L_x$ in these galaxies has a stellar origin
(Irwin \& Sarazin 1998). 
Power law ellipticals tend to be non-central in groups/clusters and 
therefore tidally subordinate with some exceptions. 
More luminous 
group-centered, group-dominant ellipticals 
typically have much larger $L_x/L_B$, 
enhanced by an additional 
contribution of circumgalactic or intragroup 
hot gas (Helsdon et al. 2001; Matsushita 2001). 
The possible influence of galactic rotation and 
(oblate) flattening on $L_x/L_B$ have been explored 
with limited success 
(Nulsen et al. 1984;
Kley \& Mathews 1995; 
Brighenti \& Mathews 1996;
Pellegrini, Held \& Ciotti 1997; 
D'Ercole \& Ciotti 1998). 
While there is some evidence that 
$L_x/L_B$ is several times lower in 
flattened and rotating ellipticals, this cannot explain 
the much larger scatter observed. 
In the evolutionary scheme of Ciotti et al. (1991) 
the scatter results from a transition from early 
Type Ia driven winds to cooling inflows, but this 
scenario (discussed below) produces too much iron in the hot gas 
(Loewenstein \& Mathews 1991).
Mathews and Brighenti (1998) showed that 
ellipticals with larger $L_x$ also have more extended 
hot gas, 
$L_x/L_B \propto (R_{ex}/R_e)^{0.60 \pm 0.20}$,
where $R_{ex}$ is the half-brightness radius for 
the X-ray image which can extend out to 
$\sim 10R_e$ or beyond. 
This correlation (see also Fukugita \& Peebles 1999
and Matsushita 2001)
may result from the tidal competition 
for diffuse baryonic gas among 
elliptical-dominated galaxy groups as they formed 
with different degrees of relative isolation. 
Ram pressure stripping must also influence $L_x/L_B$ for 
E galaxies in richer clusters (e.g. Toniazzo \& Schindler 2001). 

\section{HYDROSTATIC EQUILIBRIUM AND MASS DETERMINATIONS}

Evidence for dark halos in elliptical galaxies from stellar 
velocities has been slow in coming 
because of uncertainties in geometrical projection 
and in the anisotropy of stellar orbits as well as the 
inherent faintness of starlight beyond 
$R_e$ where dark matter may begin to dominate the potential.
Nevertheless, recent optical studies have revealed 
the presence of outwardly increasing 
mass to light ratios consistent with a dark matter 
contribution that is appreciable but 
not dominant at $\sim R_e$ (Saglia et al 1992; Carollo et al 1995;
Rix et al 1997; Gerhard et al 1998;
Emsellem et al. 1999; 
Matthias \& Gerhard 1999).
Gravitational lensing of background objects provides 
independent evidence for a dark matter component 
in elliptical galaxies and their surrounding groups 
(e.g. Keeton 2001).
Fortunately, the X-ray emitting hot gas allows 
in principle a much better 
determination of the total mass profile 
to very large radii. 
Accurate mass determinations require 
high quality X-ray observations 
of the gas density and (especially) 
temperature profiles plus reasonable assurance that the gas 
is in 
hydrostatic equilibrium and that gas pressure dominates. 
The mass distribution 
can only be determined for rather massive E galaxies, 
$L_B \gta L_{B,crit}$, 
in which the X-ray emission is dominated by gas, 
not stellar sources.

Hydrostatic equilibrium requires that 
systematic and turbulent hot gas velocities are subsonic. 
Information moving at the sound speed in the hot gas 
in E galaxies, 
$(\gamma P/\rho)^{1/2} \sim 513 T_{keV}^{1/2}$ km s$^{-1}$, 
crosses the optical half-light radius $R_e \sim 10$ kpc
in only $t_{sc} \sim 2 \times 10^7$ years. 
If the hot gas is losing energy by radiative losses, 
it should flow inward at a rate ${\dot M} \sim L_{x,bol}
/(5kT/2 \mu m_p) \approx 1.5$ $M_{\odot}$ yr$^{-1}$ 
where $L_{x,bol} \sim 5 \times 10^{41}$ erg s$^{-1}$ is 
a typical X-ray bolometric luminosity for massive E galaxies.
Using $n_e(R_e) \sim 0.01$ cm$^{-3}$ for a typical hot gas density 
at $R_e$, the systematic inflow velocity,
$u \sim {\dot M}/ n_e(R_e) m_p 4 \pi R_e^2 \sim 5$ km s$^{-1}$, 
is highly subsonic, consistent with 
hydrostatic equilibrium.
Less is known about the magnitude of turbulent motions 
in the hot gas, but 
radial velocities of diffuse optical emission lines 
from gas at $T \sim 10^4$ K in the central regions 
of E galaxies (e.g. Caon et al. 2000), 
$v_{turb} \lta 150$ km s$^{-1}$,
suggest subsonic motion, 
assuming this cold gas comoves with the local hot gas.

The condition for 
hydrostatic equilibrium $dP_{tot}/dr = -GM\rho/r^2$ 
allows a direct determination of the total mass of 
stars and dark matter within each radius:
\begin{equation}
M(r) = - rT(r) {k \over G \mu m_p}
\left( {d \log \rho \over d \log r} 
+ {d \log T \over d \log r} + {P_{nt} \over P}
{d \log P_{nt} \over d \log r} \right)
\end{equation}
where $m_p$ is the proton mass and $\mu = 0.61$ is the 
molecular weight for full ionization. 
In addition to the gas pressure $P$, an additional
non-thermal turbulent, magnetic or cosmic ray pressure $P_{nt}$ 
may be present.
In ellipticals containing strong radio sources
Faraday depolarization at radio frequencies 
provides direct evidence for $P_{nt}$ 
(e.g. Garrington et al. 1988; Garrington \& Conway 1991), 
but this pressure 
is usually ignored in most E galaxy  
mass determinations. 
Non-radiating relativistic protons may also be present
(e.g. Fabian et al. 2002a), 
so it is unclear if $P_{nt}$ can always be ignored.

The total integrated mass $M(r)$ 
can be estimated 
by using the average temperature within some radius.
Loewenstein \& White (1999) studied the ratio of the 
dimensional coefficients in the equation above, 
$\beta = \langle \sigma\rangle^2  /(k \langle T \rangle / \mu m_p)$
where $\langle \sigma\rangle^2 \propto GM(r)/r$ 
is the central stellar 
velocity dispersion (assumed to be isotropic) 
and $\langle T \rangle$ is the 
mean hot gas temperature within $6 R_e$ determined from fits to 
the thermal X-ray spectrum.  
Loewenstein \& White considered an optically complete sample 
of over 40 E galaxies (Davis \& White 1996).
Using accurate stellar mass profiles 
normalized to the fundamental plane, 
they determined that $\beta \approx 0.75 - 1.2$ 
should be expected in the absence of dark matter.
The observed values, 
$\beta \approx 0.6 \pm 0.1$, clearly require a 
dark matter component.
Both 
the gas and the dark matter are hotter than the central stars.
Loewenstein \& White conclude that dark matter 
increases from stellar values at the origin 
$\langle \Upsilon_V \rangle \equiv 
\langle M/L_V \rangle \approx  10 h_{70}$
$M_{\odot}/L_{V,\odot}$
to $\langle \Upsilon_V \rangle \approx 22 h_{70}$ 
$M_{\odot}/L_{V,\odot}$ within $6R_e$. 
Extended dark halos 
are a common property of all bright ellipticals.  

For a few bright E galaxies both  
$T(r)$ and $n_e(r)$ can be determined and 
Equation (1) can be solved directly for the total 
mass profile $M(r)$.
Figure 2a shows the electron density profile 
in NGC 4472, a well-observed massive E1 galaxy 
and the brightest galaxy in the Virgo cluster, 
at an assumed distance of d = 17 Mpc ($n_e \propto d^{-1/2}$).
The hot gas density profiles
may have small flattened cores  
but vary as $n_e \propto r^{-p}$ at larger radii 
with $p \approx 1 - 1.5$, so the gas mass 
increases outward. 
Using {\it Einstein HRI} data  
Trinchieri, Fabbiano \& Canizares (1986) showed that 
the optical and X-ray surface brightness profiles 
are almost identical for 
three bright Virgo ellipticals, NGC 4649, NGC 4636
and NGC 4472, so that $\rho_* \propto n_e^2$.
This remarkable 
result is illustrated again in 
Figure 2a where $n_e \propto \rho_*^{1/2}$ is seen to hold 
over a wide range in galactic radius.

\begin{figure}
\centerline{
\psfig{figure=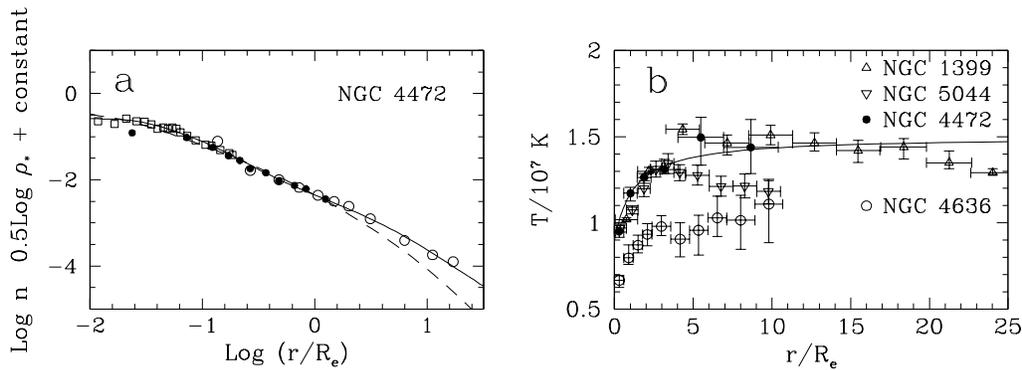,angle=270,width=7.5in} }
\caption{({\it 2a Left panel}:) The observed and
azimuthally averaged
electron density profile $n(r)$ in NGC 4472 is shown as a function
of radius normalized to the effective radius $R_e = 8.57$ kpc
at distance $d = 17$ Mpc.
The observations
are from {\it Einstein} (Trinchieri, Fabbiano,
\& Canizares 1986) ({\it filled circles}) and
ROSAT (Irwin \& Sarazin 1996) ({\it open circles});
for the inner region
we have Abel-inverted Chandra surface brightness data from
Loewenstein et al. (2001) 
({\it open squares}) and normalized them to previous
observations. 
The solid line is an analytic fit to the observations.
The dashed line is the square root of the
stellar density $\rho_*^{1/2}(r)$ normalized to $n$ 
at $r = R_e$.
({\it 2b Right panel}:) Typical temperature profiles for several
bright E galaxies, taken from Brighenti \& Mathews (1997a),
based on the following sources:
NGC 1399: ROSAT PSPC from Jones et al. (1997);
NGC 5044: ROSAT PSPC from David et al. (1994);
NGC 4636: ROSAT PSPC from Trinchieri et al. (1994);
NGC 4472: ROSAT HRI AND PSPC from Irwin \& Sarazin (1996).
The solid line is an approximate analytic fit to
$T(r)$ for NGC 4472.}
\label{fig2}
\end{figure}

Figure 2b shows the hot gas temperature profiles 
for several massive E galaxies. 
The temperature in these group-dominant E galaxies 
rises from a minimum value near 
the galactic center to a maximum at several $R_e$ and, 
if the gas is sufficiently extended, is either 
uniform or slowly decreasing beyond
(Brighenti \& Mathews 1997 and references therein), 
sometimes extending to $\gta 10 R_e$. 
(In cluster-centered E galaxies the temperature continues 
to rise to the cluster gas temperature.) 
The radiative cooling time at constant pressure in the hot gas 
in NGC 4472 is quite short, 
$t_{cool} \approx 10^8 r_{kpc}^{1.2}$ yrs, 
but greater than the dynamical time 
$t_{dyn} \approx 3 \times 10^6 r_{kpc}^{0.85}$ yrs. 
The entropy factor $T n^{-2/3}$ for NGC 4472 is 
relatively flat within $r \sim 0.55$ kpc, 
suggesting local heating (David et al. 2001), 
then increases monotonically
with radius, $T n^{-2/3} \approx 6.5 \times 10^7 r_{kpc}^{0.8712~}$
K cm$^2$, as required for convective stability.
Recent {\it Chandra} observations often show surface brightness
fluctuations and cavities, sometimes extending to $\sim R_e$,
that suggest deviations from hydrostatic equilibrium.

The total mass $M_{tot}(r)$ profile for NGC 4472 determined 
from Equation (1) (with $P_{nt} = 0$ and 
data from Figs. 2a and 2b) 
is plotted in Figure 3a. 
Also shown is the stellar mass distribution $M_*(r)$ 
based on a 
de Vaucouleurs profile $\rho_{*,deV}(r)$ 
(total mass: $M_{*t} = 7.26 \times 10^{11}$ $M_{\odot}$; 
effective radius: $R_e = 1.733' = 8.57$ kpc) 
with a core 
$\rho_{*,core}(r) = \rho_{*,deV}(r_b)(r/r_b)^{-0.90}$ 
within the break radius $r_b = 2.41'' = 200$ pc 
(Gebhardt et al. 1996; Faber et al. 1997).
It is remarkable that the total mass 
$M_{tot}(r)$ in Figure 3a determined with 
Equation (1) agrees 
quite well with the de Vaucouleurs mass profile 
in the range $0.1 \lta r/R_e \lta 1$.
The best fitting stellar profile corresponds to 
a mass to light ratio of $\Upsilon_B \equiv M/L_B \approx 7$, 
slightly less than $\Upsilon_B = 9.2$ determined for NGC 4472 from 
axisymmetric stellar models near the galactic 
core (van der Marel 1991).
This consistency of X-ray and 
stellar mass profiles suggests that the stellar mass to light ratio 
in NGC 4472 does not change greatly with galactic radius 
in $0.1 \lta r/R_e \lta 1$ (Brighenti \& Mathews 1997a; 
also for NGC 720: Buote et al. 2002a).
As X-ray observations improve we expect that they 
will provide much information on the stellar mass to light
ratio for $r \lta R_e$. 
At small radii $r \lta 0.03R_e$ in NGC 4472, 
$M_{tot}$ is less than $M_*$. 
This may indicate some additional non-thermal pressure 
$P_{nt}$ in this 
region or a deviation from hydrostatic equilibrium.
Like most bright E galaxies, NGC 4472 contains a faint 
double lobe radio source that extends 
to $\sim 0.5R_e$ (Ekers \& Kotanyi 1978).

\begin{figure}
\centerline{
\psfig{figure=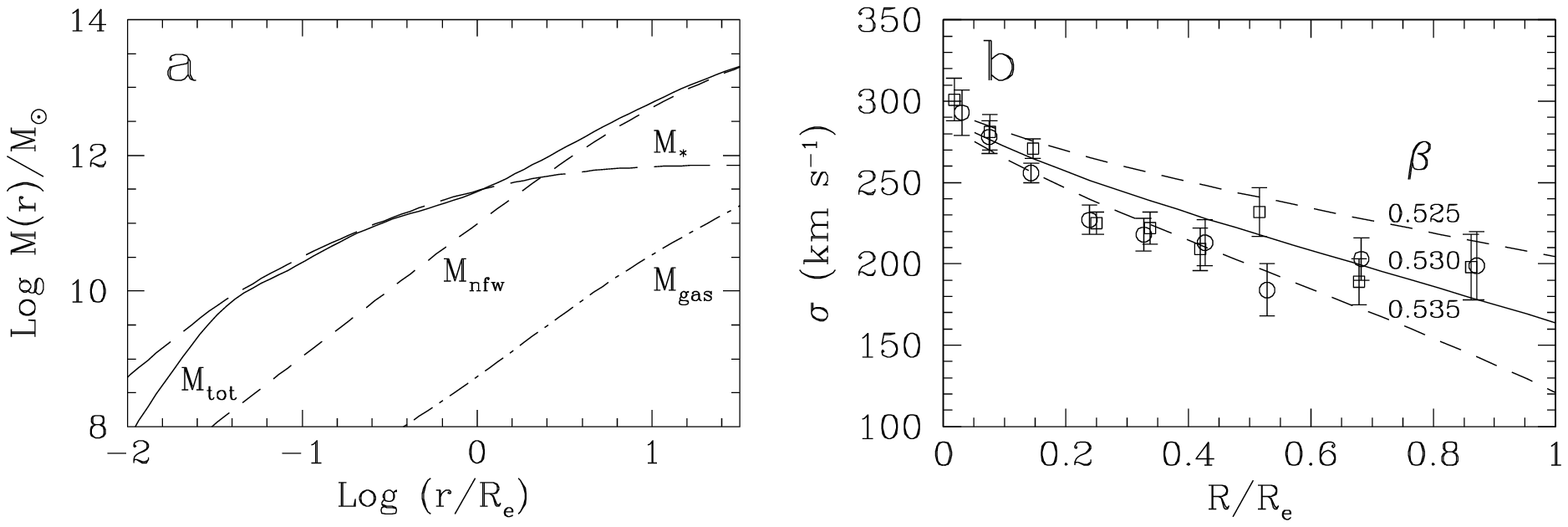,angle=270,width=7.5in} }
\caption{
({\it 3a Left panel}:) The total mass $M_{tot}(r)$
for NGC 4472 ({\it solid line}) is found from Equation (1)
with $P_{nt} = 0$ using the solid line approximations
to the X-ray observations in Figure 2.
The total mass of hot gas $M_{gas}(r)$
({\it dot-dashed line}) is relatively small.
The stellar mass profile $M_*(r)$ 
({\it long dashed line}) is based
on a de Vaucouleurs plus core profile with mass to light
ratio $\Upsilon_B = 7$. 
The NFW dark halo profile $M_{nfw}(r)$ ({\it short dashed line})
corresponds to a total mass of $4 \times 10^{13}$ $M_{\odot}$.
({\it 3b Right panel}:) Line of sight stellar velocity dispersion
profiles $\sigma(\beta; R)$ as a function of projected radius $R$.
The curves are computed from solutions $\sigma_r(r)$ of
Equation (2) assuming constant
$\beta$ which labels each curve. The observations are
from Fried \& Illingworth (1994).
}
\label{fig3}
\end{figure}

The dark halo mass 
clearly dominates in Figure 3a for $r \gta R_e$ where $M_{tot}(r)$ 
rises sharply above the de Vaucouleurs profile 
(e.g. Brighenti \& Mathews 1997a; Kronawitter et al. 2000).
The shape of the dark halo is consistent with an NFW halo 
(Navarro, Frenk \& White 1996), but the virial mass 
of the dark halo surrounding NGC 4472 and its mass profile 
are poorly determined in part due to 
uncertainties in the hot gas temperature beyond several $R_e$.
In addition, the X-ray image of NGC 4472 is asymmetric for 
$r \gta 2.5 R_e$, as seen in Figure 4, apparently because of 
its motion through the more extended Virgo cluster gas 
or possibly due to its interaction with the nearby dwarf 
irregular galaxy UGC 7636 (Irwin \& Sarazin 1996; 1997). 
(See Fabbiano et al. 1992 for an atlas of similar figures.)
In spite of these problems,
the azimuthally averaged gas density profile around NGC 4472 
is similar to the mean profile of about 10 
other bright E galaxies out to at least $18 R_e$. 

\begin{figure}
\centerline{
\psfig{figure=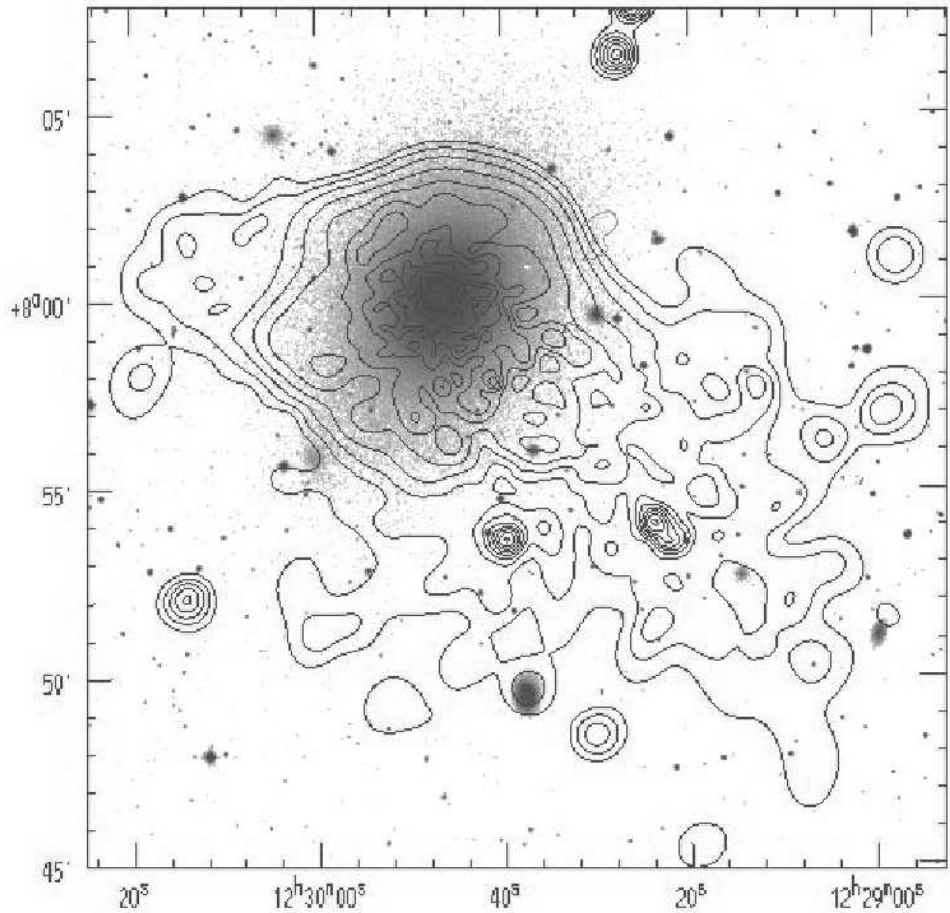,angle=0,width=5in} }
\caption{Contours show a combined
ROSAT HRI and PSPC X-ray image of NGC 4472
superimposed on an optical image from the
Digital Sky Survey (Irwin \& Sarazin 1996).}
\label{fig4}
\end{figure}

In non-spherical E galaxies, 
the existence of massive dark halos can be inferred directly 
from the X-ray image independent of the hot gas temperature 
profile -- providing the gas is in hydrostatic equilibrium 
and rotation has little influence on the potential.
For example, Buote \& Canizares (1998) find that the X-ray 
isophotes of the E4 galaxy NGC 3923 have ellipticity 
$\epsilon_x = 0.15 \pm 0.05$, which is significantly 
less than that in the R-band, $\epsilon_R = 0.30$.
Although the gravitational potential is always more spherical than 
the mass distribution,
Buote \& Canizares show that this $\epsilon_x$ 
can be understood only if the confining mass is 
greater and more extended 
than any mass distribution proportional to the optical light
(also: Buote et al. 2002a). 
Furthermore, the dark mass must have an ellipticity 
$\epsilon_{dm} = 0.35 - 0.66$ greater than the optical image.  
Not only is this an elegant method to detect dark matter
(see also Buote \& Canizares 1996; 1997), 
it serves as a warning, generally ignored, that the dark 
matter may not be distributed in a quasi-spherical fashion. 

While the hot gas in NGC 4472, NGC 4649 and NGC 720  
appears to
be in hydrostatic equilibrium in the {\it stellar} potential
in $0.1 \lta r/R_e \lta 1$, this 
circumstance may not be universal.
For example, in NGC 4636 Brighenti \& Mathews (1997a) found that
the total mass $M(r)$ profile found from Equation (1)
(with $P_{nt} = 0$) intersects the stellar
mass $M_*(r)$ (using $\Upsilon_B = 10.7$) with no slope change
whatsoever.
Brighenti \& Mathews argued that this insensitivity may
be due to a magnetic field, $B \sim 100\mu$G,
($P_{nt} \approx B^2/8\pi$) at $r \sim 0.1R_e$,
so that the missing $(P_{nt} / P) (d \log P_{nt} / d \log P)$
term in Equation (1) would account for the discrepancy.
Cosmic rays or energetic turbulence would serve equally well. 
In fact the Chandra X-ray image of NGC 4636 shows that the 
hot gas is strongly agitated for $r \lta R_e$ 
(Jones et al. 2002), consistent with a breakdown in 
hydrostatic equilibrium. 
Alternatively, 
in reconciling the total mass $M(r)$ 
of NGC 4636 from X-ray data, 
Loewenstein \& Mushotzky (2002) 
reduced the contribution of the stars by 
lowering the stellar mass to light ratio to 
$\Upsilon_B < 5.4$ (at $d = 17$ Mpc)
which is very much less than values determined 
for NGC 4636 from stellar velocities: 
$\Upsilon_B = 10.7$ (van der Marel 1991)
or $\Upsilon_B = 11.3$ (Kronawitter et al. 2000) 
(both at $d = 17$ Mpc). 


The shape of the stellar velocity 
ellipsoid $\beta(r)$ can also be estimated directly from X-ray 
observations, especially for very massive E galaxies 
that are approximately spherical. 
Here $\beta(r) = 1 - \sigma_t^2/\sigma_r^2$ depends on 
$\sigma_r$ and $\sigma_t = \sigma_{\theta} = \sigma_{\phi}$, 
the velocity dispersions in the radial and transverse
directions.
Combining the Jeans Equation for the radial stellar velocity 
dispersion $\sigma_r$ with the equation for hydrostatic 
equilibrium in the gas we find 
\begin{equation}
{d \sigma_r^2 \over d r} + {\sigma_r^2 \over r}
\left[ {d \log \rho_* \over d \log r} + 2 \beta \right]
= {d c^2 \over d r} 
+ \left[{c^2 \over r} {d \log \rho \over d \log r} \right]
\end{equation}
where $c^2 = kT / \mu m_p$ is the isothermal sound speed. 
From Figure 2a for NGC 4472 we see that 
$0.5(d \log \rho_* / d \log r) = d \log \rho / d \log r = -1.18$ 
fits over $-1.3 \lta \log(r/R_e) \lta 0$. 
The gas temperature variation is approximately linear 
over this region, 
$c^2(r) \approx 4135(r_{kpc} + 30)$ (km s$^{-1}$)$^2$ (Figure 2b).
If $\beta$ is assumed to be constant,
Equation (2) can be solved analytically for 
$\sigma_r^2(r)$ 
and the line of sight stellar velocity 
dispersion as a function of projected radius $\sigma (\beta ; R)$
can be found by integration (e.g. Binney \& Mammon 1982).
The resulting $\sigma(\beta;R)$, 
when compared with stellar velocity dispersion observations 
in Figure 3b, suggests $\beta = 0.530 \pm 0.005$, 
somewhat higher than that of Kronawitter et al. (2002) who 
use different velocity data. 
Conversely, if $\beta(r)$ is known securely from stellar data,
Equation (2) can be used to determine the gas temperature 
profile $c^2(r)$. 

\section{GAS DYNAMICAL EQUATIONS AND CLASSIC COOLING FLOWS}

To understand cooling flows at the next 
level beyond static models,   
the assumption of steady state flow is often made 
(Bailey 1980;
White \& Chevalier 1983, 1984; 
Nulsen, Stewart \& Fabian 1984;
Thomas 1986;
Sarazin \& White 1987; 
Vedder, Trester \& Canizares 1988;
Sarazin \& Ashe 1989;
Tabor \& Binney 1993;
Bertin \& Toniazzo 1995).
Although the steady state approximation is useful in gaining 
insight into various dynamical aspects of subsonic cooling flows, 
particularly at small galactic radii, 
inconsistencies can arise if this approximation 
is applied globally. 
In particular for subsonic flows there 
are ambiguities in selecting the boundary conditions near the 
stagnation radius where the inward integrations
begin, as  
recognized and discussed by Vedder, Trester \& Canizares (1988). 
Steady flows are also incapable of properly allowing 
for some essential time dependencies such as the 
strongly decreasing rate of stellar mass loss, 
variations in the frequencies of 
Type Ia supernovae, heating by central AGN, and the 
accretion of intergalactic gas in an evolving cosmology.
For these reasons we emphasize non-steady 
flows in the following discussion.

For galactic scale cooling flows, the usual 
time-dependent gas dynamical 
equations must include appropriate source and sink terms:
\begin{equation}
{ \partial \rho \over \partial t}
+ {\bf \nabla} \cdot \rho {\bf u}
= \alpha \rho_*
\end{equation}
\begin{equation}
\rho { d {\bf u} \over d t} = 
\rho \left[ { \partial {\bf u} \over \partial t}
+ ({\bf u} \cdot {\bf \nabla}){\bf u} \right]
= - {\bf \nabla} P 
- \rho {\bf \nabla}\Phi - \alpha \rho_* ({\bf u} - {\bf u}_*),
\end{equation}
and
\begin{equation}
\rho {d \varepsilon \over d t} 
- {P \over \rho} {d \rho \over dt} = 
- { \rho^2 \Lambda \over m_p^2}
+ \alpha \rho_*
\left[ \varepsilon_o - {P \over \rho} - \varepsilon 
+ {1 \over 2}|{\bf u} - {\bf u}_*|^2  \right]
- {\bf \nabla} \cdot {\bf F}_{cond}
+ H(r,t) \rho
\end{equation}
where $\varepsilon = 3 k T / 2 \mu m_p$ is the
specific thermal energy. 
The gravitational potential $\Phi$ in the momentum 
Equation (4) 
includes contributions from both stellar and dark mass;
the mass of hot gas is usually negligible.
The coefficient $\alpha = \alpha_* + \alpha_{sn}$ is 
the specific rate of mass loss from the stars and Type Ia 
supernovae, i.e. $\alpha M_{*t} 
\approx 0.15 M_{*t}/(10^{11} M_{\odot})$ 
$M_{\odot}$ yr$^{-1}$ is the total current rate of mass loss 
from evolving stars in NGC 4472.
Because $\alpha_{sn}$ is much less than 
$\alpha_*$, $\alpha \approx \alpha_*$ 
is an excellent approximation. 
The specific mass loss rate from 
a single burst stellar population with Salpeter IMF 
and age $t$ varies as 
$\alpha_*(t) = 4.7 \times 10^{-20} (t/t_n)^{-1.3}$ s$^{-1}$ 
where $t_n = 13$ Gyrs (Mathews 1989); 
for other powerlaw IMFs $\alpha_*(t_n)$ varies inversely 
with the stellar mass to light ratio.

A fundamental assumption is that gas ejected from 
evolving giant stars
as winds or planetary nebulae eventually becomes 
part of the hot phase.
The interaction of gas ejected from orbiting stars with 
the hot galactic gas is extremely complicated, involving  
complex hydrodynamic instabilities that enormously increase 
the surface area between the ejected gas and its hot environment 
until, as usually assumed, the two gases thermally fuse. 
As we discuss below, the positive temperature gradients 
observed in inner cooling flows of cluster-centered 
elliptical galaxies and the hot gas oxygen abundance 
gradients provide indirect evidence 
that such mixing does occur. 
The term $\alpha \rho_* ({\bf u} - {\bf u}_*)$ 
in Equation (4) represents a  
drag on the flow, assuming gas expelled from stars has on average 
the mean stellar velocity ${\bf u}_*$, 
which is non-zero only in rotating galaxies. 
This drag term is generally negligible if the flow is subsonic.  
If $u_* = 0$, 
this term has the effect of making subsonic flow even more 
subsonic and supersonic flow even more supersonic.

The thermal energy Equation (5) contains a term 
$-(\rho/m_p)^2 \Lambda(T,z)$ for the loss of energy by 
X-ray emission; the radiative cooling coefficient 
$\Lambda(T,z)$ erg cm$^3$ s$^{-1}$ 
varies with both gas temperature and metal 
abundance (e.g. Sutherland \& Dopita 1993). 
The dissipative heating $\alpha_*\rho_* |{\bf u}-{\bf u}_*|^2/2$
involved in accelerating stellar
ejecta to the local flow velocity is usually very small.
The hot gas temperature is also influenced by stellar mass 
loss and Type Ia supernovae. 
The source terms 
$\alpha_*\rho_* (\varepsilon_o - P /\rho - \varepsilon)$ 
represent the heating of the hot interstellar gas 
of specific energy $\varepsilon$ by the mean energy 
of stellar ejecta $\varepsilon_o$ less the work done 
$P/\rho$ in displacing the hot gas.
The mean gas injection energy is
$\varepsilon_o = 3 k T_o /2 \mu m_p$ where
$T_o = (\alpha_* T_* + \alpha_{sn} T_{sn})/\alpha$.
The stellar temperature $T_*$ can be found by
solving the Jeans equation, 
but this term is small and it is often sufficient to use 
an isothermal approximation, $T_* = (\mu m_p/k)\sigma^2$, 
where $\sigma$ is the average stellar velocity dispersion.
Supernova heating is assumed to be distributed
smoothly in the gas, ignoring the detailed 
evolution of individual blast waves (Mathews 1990).
The heating by Type Ia supernovae,
each of energy 
$E_{sn} \approx 10^{51}$ ergs, is described by multiplying
the characteristic temperature of the
mass $M_{sn}$ ejected, 
$T_{sn} = 2 \mu m_p E_{sn} / 3 k M_{sn}$,
by the specific mass loss rate from supernovae, 
$\alpha_{sn} = 3.17 \times 10^{-20} {\rm SNu}(t) 
(M_{sn}/M_{\odot}) \Upsilon_B^{-1}~~~{\rm s}^{-1}$.
Here the supernova rate SNu is expressed in 
the usual SNu-units,
the number of supernovae in 100 yrs expected from stars of total
luminosity $10^{10}L_{B\odot}$.

Supernovae in ellipticals today are infrequent and all of 
Type Ia.
Cappellaro et al. (1999) find 
SNu$(t_n) = (0.16 \pm 0.05)h_{70}^2$ for E+S0 galaxies.
The past evolution of this rate SNu$(t)$
is unknown, although some provisional data is beginning 
to emerge (Gal-Yam, Maoz, \& Sharon 2002).
Clearly, it would be very useful to have more information about
SNu$(t)$ for E galaxies 
at high redshift because this can have a decisive 
influence on the evolution of the hot gas and its iron abundance. 
Type Ia supernovae may involve mass exchange between 
binary stars of intermediate mass, but the details 
are very uncertain (e.g. Hillebrandt et al. 2000). 
However, like all cosmic phenomena, it is generally assumed that 
SNu$(t)$ is a decreasing function of time,
${\rm SNu}(t) = {\rm SNu}(t_n)(t/t_n)^{-s}$.
This is consistent with measurements at intermediate 
redshifts if $s \sim 1$ (Pain et al. 2002), although 
these observations refer to all galaxy types.

Ciotti et al. (1991) recognized that
the relative rates of stellar mass ejection 
($\alpha_* \sim t^{-1.3}$)
and Type Ia supernova (SNu $\sim t^{-s}$)
determine the dynamical
history of the hot interstellar gas in ellipticals.
For example, if
$s > 1.3$ in isolated ellipticals 
then the supernova energy per unit mass of 
gas expelled from stars ($\propto \alpha_{sn}/\alpha_*$) 
was large in the distant past, 
promoting early galactic winds, 
but if $s < 1.3$, outflows or 
winds tend to develop at late times.
However, in the presence of circumgalactic gas, 
the early time 
galactic winds driven by Type Ia supernovae can be suppressed. 
Each Type Ia supernova injects $\sim 0.7$ $M_{\odot}$
of iron into the hot gas, producing a
negative iron abundance gradient 
in the hot gas that depends on  
$\rho_*/\rho$ and the radial flow velocity of the gas. 
The observed SNIa enrichment provides an 
important constraint on $d$SNu$(t)/dt$ 
(Loewenstein \& Mathews 1991).
Evolutionary flow solutions with $s > 1.3$ produce 
iron abundances far in excess of those observed today, 
unless the iron is preferentially 
removed by selective cooling. 
We therefore have tentatively adopted $s = 1$ with 
${\rm SNu}(t_n) = 0.06$ SNu 
(similar to the estimate of 
Kobayashi et al 2000) although this is by no means 
the only possibility. 
This adopted current Type Ia supernova rate 
is less than the rate observed 
in E plus S0 galaxies, but the rate 
may be lower for ellipticals than for S0 galaxies.

Thermal conductivity in a hot plasma,
$\kappa \approx 5.36 \times 10^{-7} T^{5/2}$ 
erg s$^{-1}$ cm$^{-1}$ K$^{-1}$ is important at 
high temperatures, but may be reduced by 
tangled magnetic fields. 
The conductive energy flux in Equation (5), 
$F_{cond} = f\kappa dT/dr$, usually includes an 
additional factor $f \le 1$ to account for magnetic 
suppression.
In the past $f \ll 1$ has often been assumed, but
Narayan \& Medvedev (2001) have recently shown that
$f \sim 0.2$ is appropriate for
thermal conduction in a hot plasma with
chaotic magnetic field fluctuations. 
The final term in Equation (5), $H(r,t)\rho$, 
is an {\it ad hoc} AGN heating term that is discussed 
later.

Using the relation $\rho_* \approx 8.54 \times 10^{-20} 
n_e^2$ from Figure 2a, we find at the current time $t_n$ that 
$(\rho/m_p)^2 \Lambda$ is about an order of magnitude greater 
than $\alpha_*\rho_*(\varepsilon_o - P/\rho - \varepsilon)$. 
For galactic flows with $T \sim 10^7$ K
thermal conduction is important for $f \gta 0.5$.
Therefore, if dissipation and AGN heating are small 
in NGC 4472, 
$(\rho/m_p)^2 \Lambda$ dominates all other 
non-adiabatic terms in Equation (5), 
generating a classic cooling inflow driven by radiative losses.
As radiative energy is lost 
in a Lagrangian frame moving with the gas,
the entropy decreases, but the gas temperature 
$\sim T_{vir}$ remains relatively constant as the gas 
is heated in the gravitational potential by $Pdv$ compression.
The compression drives gas slowly toward the galactic center 
where, in this simple example, $Pdv$ heating is no longer 
available and catastrophic cooling ensues.
Because of this self-regulating mechanism, 
the temperature profile 
$T(r)$ in cooling flows 
is very insensitive to modest changes in 
the source terms in the thermal energy Equation (5), 
including $(\rho/m_p)^2 \Lambda$.

To gain further insight, it is instructive to insert 
the observed gas density and temperature 
profiles for NGC 4472 into Equations (3) and (5) 
and estimate the steady state 
radial gas velocity, assuming $\alpha_*$ 
is fixed at its current value.
For a steady inflow Equation (3) can be integrated from $r$ to 
$\infty$ and solved for the flow velocity,
$u_{\alpha}(r) = \{ {\dot M}(\infty) 
- \alpha_*[M_{*t} - M_*(r)] \}/4 \pi r^2 \rho$.
This is the negative velocity required to continuously 
remove mass supplied by mass-losing galactic stars
without changing $\rho(r)$. 
Assuming ${\dot M}(\infty) \approx 0$ and 
using NGC 4472 parameters, we find 
$u_{\alpha}(r) \approx - 51 r_{kpc}^{-1.05}$
km s$^{-1}$ for $0.25 \lta r_{kpc} \lta 10$; at larger $r$ in 
NGC 4472 the inflow of circumgalactic gas ${\dot M}(\infty) < 0$
must be considered but for $r \lta R_e = 8.57$ kpc 
we can assume the mass lost from the stars determines $\rho(r)$. 
Another steady state velocity can be found by inserting the 
observed $\rho(r)$ and $T(r)$ (Fig. 2) into  
the first three terms of Equation (5),
\begin{equation}
u_{\Lambda}(r) = -\left( {3 \over 2} {d \log T \over d \log r}
- {d \log \rho \over d \log r} \right)^{-1} {\mu \over k}
{\rho \over m_p} { r \Lambda(T) \over T}.
\end{equation}
This ``slump'' velocity in NGC 4472, 
$u_{\Lambda}(r) \approx -27 r_{kpc}^{-0.36}$ km s$^{-1}$ 
for $0.3 \lta r_{kpc} \lta 40$, 
is the inflow that occurs because 
gas is cooling near the center and occupying less volume.
Both $u_{\alpha}$ and $u_{\Lambda}$ are very small compared to 
the adiabatic sound speed in the hot gas, 
$476 (T/10^7~{\rm K})^{1/2}$ 
km s$^{-1}$, so either type of flow satisfies the 
requirement for hydrostatic equilibrium. 
However, for $r_{kpc} \lta 2.5$, $|u_{\alpha}| > |u_{\Lambda}|$,
i.e., the inflow required to conserve the observed 
gas density profile exceeds the rate that the gas can cool,
violating the assumption of steady flow.
Consequently, if the observed $\rho(r)$ and $T(r)$ are taken 
as initial conditions in a time-dependent gasdynamical 
calculation for NGC 4472, 
subsonic inflowing 
solutions evolve toward higher gas densities near the origin, 
increasing the radiative losses there 
until $u_{\alpha}(r) \approx u_{\Lambda}(r)$. 
This explains why the central gas density exceeds observed 
values in every otherwise successful steady state or 
time-dependent inflow model without 
central AGN heating or thermal conduction.

\section{SIMPLE COOLING FLOW MODELS}

To illustrate the influence and relevance 
of various terms in Equations (3) - (5), we briefly 
describe several simple time-dependent solutions 
(also see Loewenstein \& Mathews 1987). 
For a reference flow we consider an E0 galaxy with
no conductive or AGN heating
and no source of gas except that lost from the stars, 
i.e. an ``isolated'' elliptical galaxy.   
The calculations begin at 
cosmic time $t_{in} = 1$ Gyr when we imagine 
that the (recently assembled) galaxy has just 
been cleared of gas by Type II supernovae.
We assume NGC 4472 parameters with an NFW group halo 
of mass $M_h = 4 \times 10^{13}$ $M_{\odot}$ and 
supernova rate 
${\rm SNu}(t) = {\rm SNu}(t_n)(t/t_n)^{-s}$ with 
${\rm SNu}(t_n) = 0.06$ and $s = 1$. 
In this reference model gas cools only at the galactic 
center and for simplicity we ignore the (not insignificant) 
gravitational influence of the cooled gas on 
hot gas near the origin; with this assumption the 
solutions are less sensitive to $t_{in}$. 
After the flow evolves to time $t_n = 13$ Gyrs 
we compare the density and temperature profiles with 
those of NGC 4472 and consider the mass that has 
cooled and the iron abundance in the hot gas.
The stellar iron abundance is 
$z_{*,Fe} = 0.75[1 + (r/R_e)^2]^{-0.2}$ (in solar 
meteoritic units) 
and 0.7 $M_{\odot}$ of iron is contributed by each 
Type Ia supernova.
Then we describe the effect on the reference solution when 
one of the many terms and 
parameters in Equations (3) - (5) is altered. 
None of these models agrees completely 
with the observations although some agree 
much better than others.

{\bf Reference Flow:} The reference model at time $t_n$, 
shown with solid lines in column 1 of Figure 5, 
clearly has a steeper density slope than the observed profile 
and the temperature beyond about 5 kpc is too low.
Both of these discrepancies occur in large part because 
our reference model galaxy is ``isolated''. 
The real NGC 4472 galaxy is (or was) surrounded by an 
extended circumgalactic cloud of gas 
at the somewhat higher virial temperature of the 
dark matter 
potential of the galaxy group from which NGC 4472 formed. 
As we discussed earlier,
the excess gas density inside $r \sim 10$ kpc is 
a characteristic feature of all flow models except those 
with additional non-thermal pressure or heating.
The temperature gradient 
$dT/dr$ is negative in disagreement with observations 
of NGC 4472 and other similar galaxies (Figure 2b).
Negative $dT/dr$ occurs because of the steepness 
of the stellar potential in $r \lta R_e$. 
When the reference calculation 
is repeated ignoring the stellar gravity but retaining 
only the softer NFW potential, $T(r)$ passes through a maximum 
around $r \sim 40$ kpc and $dT/dr > 0$ within this radius, 
similar to cooling flow thermal profiles in rich clusters. 
For the reference flow we find 
$L_{x,bol}(t_n) = 6.7 \times 10^{41}$ ergs s$^{-1}$ 
from the hot diffuse gas. 
The iron abundance in the hot gas at $r \sim 10$ kpc,  
$z_{Fe}/z_{Fe,\odot} \sim 2.7$ (meteoritic), is 
slightly higher than
observed in the hot gas of NGC 4472, 
$z_{Fe}/z_{Fe,\odot} \sim 2$ 
(e.g. Buote et al. 2000a, 2000c), 
even though our reference SNu$(t_n) = 0.06$ is rather low. 

\begin{figure}
\centerline{
\psfig{figure=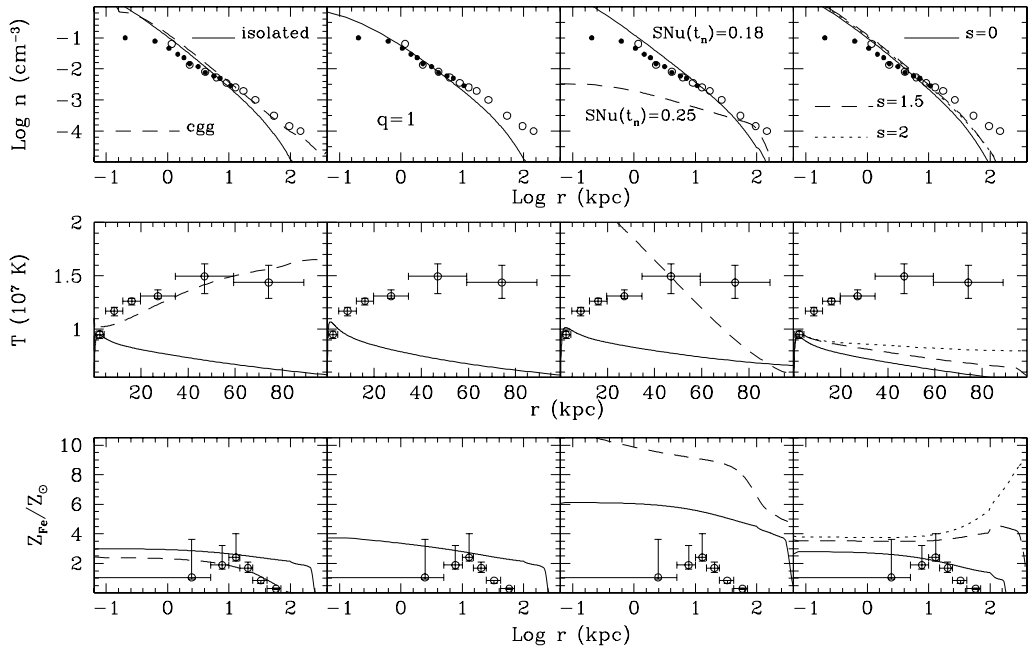,width=8in,angle=270} }
\caption{A variety of computed
time dependent galactic flow models
are compared at time $t_n = 13$ Gyrs
with the observed gas density ({\it top row})
and temperature ({\it middle row}) of NGC 4472 taken
from Figure 2.
The bottom row shows the computed hot gas iron abundance
compared with the observations of Buote (2000c).
{\it First column:} A plot of the profiles at $t_n$ for
an approximate model of NGC 4472 assuming an isolated
galaxy ({\it solid lines}); this ``reference'' flow
is based on an assumed parameter set as explained in the
text. A similar galactic flow but including 
initial circumgalactic gas (``cgg'') is
also shown ({\it dashed lines}).
{\it Second column:} A model similar to the reference
flow but with {\it ad hoc} spatially distributed
cooling with dropout parameter $q = 1$.
{\it Third column:} Two flows similar to the reference
flow (for which SNu$(t_n) = 0.06$) but with additional heating
by Type Ia supernovae: SNu$(t_n) = 0.18$ ({\it solid lines})
and SNu$(t_n) = 0.25$ ({\it dashed lines}).
{\it Fourth column:} Three galactic flows similar
to the reference flow but calculated
with different past Type Ia supernova rates 
SNu$(t) \propto t^{-s}$: $s = 0$ ({\it solid lines}),
1.5 ({\it dashed lines}) and 2 ({\it dotted lines}).
In the reference model $s = 1.3$.}
\label{fig5}
\end{figure}

{\bf Circumgalactic Gas:} 
The dashed lines in column 1 of Figure 5 show the 
effect of including circumgalactic, group-related gas 
in the NGC 4472 evolution 
(Thomas 1986; Bertin \& Toniazzo 1995; 
Brighenti \& Mathews 1998). 
Agreement with the observations is improved in several ways: 
(1) By design 
the gas density beyond the optical galaxy ($r \gta 10$ kpc) 
is increased to fit the data. 
(2) Since the virial temperature of the NFW group halo 
exceeds that of the central galaxy, the gas temperature 
of the circumgalactic gas is higher.
As hot circumgalactic gas flows inward,
it is cooled by radiation losses 
and by mixing with stellar ejecta at characteristic temperature 
$\sim T_* \sim 10^7$ K, naturally 
creating the observed positive temperature gradient 
within $\sim 50$ kpc. 
All known E galaxy temperature profiles 
are positive (Figure 2b) within several $R_e$ and 
must therefore contain hot circumstellar gas, 
but when those with very small X-ray images are observed 
(e.g. NGC 4374: Mathews \& Brighenti 1998),  
it is possible that $dT/dr$ will be negative as in the 
``isolated'' reference model. 
(3) If the iron abundance in the inflowing circumgalactic 
gas is $z_{Fe}/z_{Fe,\odot} \sim 0.3 - 0.4$, typical of external 
regions in groups and clusters, then as the inflowing gas 
mixes with supernova ejecta, the mean 
abundance $z_{Fe}/z_{Fe,\odot} \sim 1 - 2$ 
at $r \sim 10$ kpc 
is closer to observed values.
The beneficial effect of these improvements, taken together, 
provide support for subsonic inflow as 
in classical cooling flow theory. 

{\bf Central Mass Difficulties:}
Our reference solution
has serious problems near the origin. 
The amount of gas that has cooled there by time $t_n$, 
$M_{cold} = 37 \times 10^{9}$ $M_{\odot}$, far exceeds the mass 
$M_{bh} \sim 0.56 - 2.6 \times 10^9$ $M_{\odot}$ of the 
central black hole observed in NGC 4472
(e.g. Magorrian et al. 1998; Merritt \& Ferrarese 2001) 
and would cause the 
central stellar velocity dispersion to exceed 
the observed value. 
In standard cooling flows in luminous E galaxies, 
with or without circumgalactic gas, 
the mass of cooled gas, $M_{cold}$, 
is several times greater than the
total mass of hot gas at $t_n$.
Although most of the cooled mass $M_{cold}$ is
formed at early times ($\alpha_* \propto t^{-1.3}$)
when our understanding of galactic evolution is uncertain, 
at the current cooling rate for the reference flow,
${\dot M} \approx 1.1$ $M_{\odot}$ yr$^{-1}$,
$M_{cold}$ and $M_{bh}$ become equal after only $\sim 2$ Gyrs.
$M_{cold}$ can also be reduced
by supernova driven galactic winds at early times;
this may occur even before the E galaxy formed by mergers.
If the reference model is begun gas-free at 
a much later time
$t_{in} = 5$ Gyrs (redshift $z \sim 1.25$ for $H_0 = 70$,
$\Omega_m = 0.3$ and $\Omega_{\Lambda} = 0.7$), then 
$M_{cold} = 7.7 \times 10^{9}$ $M_{\odot}$ is reduced 
but still exceeds 
$M_{bh}$ -- the problem of excessive $M_{cool}$ 
does not go away easily. 

Compact, luminous X-ray emission is expected as
hot interstellar gas flows toward the central supermassive
black holes in E galaxies (Fabian \& Canizares 1988).
In our reference flow, even if the potential 
energy of the black hole is ignored, 
the X-ray luminosity of gas cooling
by thermal emission at the very center of the
flow, $L_{x,bol}(r=0,t_n) \approx (5kT/2\mu m_p){\dot M}(0)
\approx 10^{41} T_7 ({\dot M}/M_{\odot}~{\rm yr}) 
\sim 10^{41}$ ergs s$^{-1}$,
is comparable to $L_{x,bol}$ from the rest of the flow,
in flagrant violation of observations. 
Attempts to detect compact X-ray sources in giant
E galaxies have been remarkably unsuccessful
(e.g. Fabian \& Rees 1995;
Reynolds et al. 1996; 
Di Matteo et al. 2000;
Roberts \& Warwick 2000; 
Loewenstein et al. 2001; 
Sulkanen \& Bregman 2001).
This emission shortfall is usually expressed 
in terms of the luminosity 
$L \approx \eta {\dot M} c^2$ of 
spherical Bondi accretion onto a mass point,
${\dot M} \approx 4 \pi \rho_{\infty} c_{\infty}^{-3}
(GM_{bh})^2$,
where $\rho_{\infty}$ is the density of distant gas at rest 
and $c_{\infty} \approx 370 T_7^{1/2}$ km s$^{-1}$ 
is the isothermal sound speed in this gas.
If energy is produced with efficiency 
$\eta = 0.1$, the expected luminosity 
$L \sim \eta {\dot M} c^2 \sim 6 \times 10^{44} (\eta/0.1)
n_e T_7^{-3/2} (M_{bh}/10^9~M_{\odot})^2$, is similar to a quasar.
Central X-ray emission from 
Chandra observations of NGC 6166 (Di Matteo et al. 2001)
and M87 (Di Matteo et al. 2003), where nuclear 
X-ray sources are observed, 
indicate $\eta \sim 10^{-4}$. 
This low efficiency is within 
the radiation-reducing capability of 
advection dominated accretion flows (ADAFs) 
(Rees et al. 1982; Narayan \& Yi 1995; Abramowicz et al. 1995).
The ${\dot M}$ in M87 and NGC 6166 may also be reduced 
below the Bondi rate by occasional AGN heating.
In addition, some of the accreting mass and energy may  
be redirected to kinetic flow along 
a jet (Blandford \& Begelman 1999) which for M87 
is $\sim 10^{44}$ erg s$^{-1}$. 
Recently Loewenstein et al (2001) examined {\it Chandra}
images of several bright E galaxies
(NGC 1399, NGC 4472, and NGC 4636) and found no evidence
of compact nuclear X-ray emission in the galactic cores, 
indicating $\eta \lta 10^{-5}$. 
Either the radiative efficiency is incredibly low or gas 
is not arriving at the black hole in Bondi flow.
Perhaps the gas is outflowing in this 
region or heated in some way by the black hole;  
these possibilities 
would be compatible with the flat density gradient observed 
in the central few kpc of M87 (Di Matteo et al. 2002) 
that is difficult to produce with inflowing or static solutions.

{\bf Mass Dropout:}
The traditional device 
to avoid huge central masses of cooled gas 
and central inflows at $t_n$ 
has been to assume that the cooling is somehow spread over a large  
range of radius.
To accomplish this, 
an {\it ad hoc} ``dropout'' term is added 
to the right hand side of 
Equation (3), $-q \rho/t_{cool}$, 
where $q$ is a dimensionless parameter
(e.g. Fabian, Nulsen \& Canizares 1984a;
White \& Sarazin 1987; Sarazin \& Ashe 1989; Kritsuk 1992).
This term is designed to force the 
gas to cool at any radius in 
proportion to the local gas density divided by the local 
cooling time at constant pressure,
$t_{cool} =  5 m_p k T/ 2 \mu \rho \Lambda$.
For constant $q$ 
the rate of cooling dropout, $q \rho/t_{cool} \propto n_e^2$, 
is concentrated toward the galactic center, 
but $q$ can also be assumed to vary with galactic radius.
Sarazin \& Ashe (1989) showed that models with $q \approx 1$ 
fit the X-ray data reasonably well. 
It is interesting to estimate the value of $q$ 
that just balances mass loss from stars in Equation (3),
$q = (t_{cool}/\rho)\alpha_* \rho_* 
\approx 0.4$ where $T = 10^7$ K is assumed 
and we use $\rho_* \approx 8.54 \times 10^{-20}
n_e^2$ from Figure 2a.
Therefore, in evolutionary models with uniform $q \sim 1$ 
gas is removed from the flow at about the same rate that 
it is supplied by stars. 
The flow does not shut down if $q > 1$, however, since 
the gas density decreases and $q \rho/t_{cool} \propto \rho^2$
becomes less effective; the total mass of cooled 
gas is quite independent of $q$. 

In column 2 of Figure 5 we plot the density and temperature  
using reference model parameters but including the dropout term
in Equation (3) with $q = 1$. 
One of the historical motivations for dropout was to  
avoid the central rise in gas density in 
cooling flow models, but in 
our experience dropout does not completely solve this difficulty 
as seen in Figure 5. 
In this dropout solution the mass of cooled gas, 
$M_{cold} = 3.8 \times 10^{10}$ $M_{\odot}$, 
is almost the same as in the reference solution, but 
only a small fraction cools at the origin. 
In flows with mass dropout the gas is multiphase 
everywhere, 
i.e. some gas cools in pressure equilibrium at every radius 
and passes through a continuum of higher densities and 
lower temperatures. 
The additional emission from these cooling regions, 
if they exist, 
is substantial and must be added to the emission of 
the smooth background gas that radiates in the normal way.
Consequently, the observed or apparent gas density is higher 
and the temperature lower than that of the smooth background. 
The apparent density profile $n_e(r)$ in column 2 of Figure 5 
agrees much better with the data for NGC 4472 than 
the reference model ($q = 0$) in column 1. 
Adding circumgalactic gas would improve the 
agreement further.
 
If the cooled gas forms into a spatially extended population
of optically dark (dwarf) stars, as often assumed,
then the stellar mass to light ratio would vary with
galactic radius. 
In some cases this dark mass can thicken 
and distort the fundamental plane beyond observed limits 
(Mathews \& Brighenti 2000).
But in the high pressure environment of galactic flows, 
stable Bonner-Ebert spheres at $10^4$ K have masses 
$\lta 2 M_{\odot}$, and this may also be the maximum 
mass of any stars that form  
(Mathews \& Brighenti 1999a); because these stars are 
optically luminous, their influence on the
fundamental plane is lessened.
Young stars in this mass range could explain the 
high stellar H$\beta$ features that are commonly observed 
in giant E galaxies 
(Mathews \& Brighenti 1999b; Terlevich \& Forbes 2002).

Another historic difficulty with the 
dropout hypothesis is that infinitesimal perturbations 
in the gas density do not develop into full blown 
thermal instabilities (e.g. Balbus 1991).
Loewenstein (1989) showed that 
small (coherent!) density perturbations oscillate 
radially in the nearly static hot gas atmosphere 
with very little overdensity on average and do 
not cool appreciably faster than the ambient undisturbed gas. 
Computational studies of the gas dynamics of initially
overdense regions in cooling flows 
(Hattori \& Habe 1990;
Yoshida, Habe \& Hattori 1991;
Malagoli, Rosner \& Fryxell 1990; Reale et al. 1991;
Hattori, Yoshida \& Habe 1995)
indicate that runaway thermal instabilities are not expected 
unless the initial perturbation
amplitude is very large, $\delta \rho / \rho \gta 1$.
However, in recent 2D calculations of AGN heated 
flows spatially distributed cooling appeared spontaneously 
near the outer boundary of the convective region 
(Kritsuk, Plewa \& M\"uller 2001) and also 
in non-linear compressions in  
convective regions 
(Brighenti \& Mathews 2002b).

Nevertheless, intermediate (multiphase) temperatures, 
an essential outcome of radiative cooling and mass dropout, 
are not supported by XMM X-ray spectra 
of galactic scale flows 
(NGC 4636: Xu et al. 2002; 
NGC 5044: Buote et al. 2003a;
NGC 1399: Buote 2002; 
M87: Molendi \& Pizzolato 2001).
Likewise, in cluster scale flows there is no evidence for gas
cooling below $\sim 1 - 2$ keV 
(Peterson et al. 2001; Tamura et al. 2001;
Kaastra et al. 2001; Molendi \& Pizzolato 2001;
B\"ohringer et al. 2002; Matsushita et al. 2002).
These astonishing null results have led to many 
speculations, discussed below, 
but at present no single explanation is generally accepted. 

{\bf Transition to Winds:}
Clearly, it could be helpful 
if the gas flowed out rather than in, 
but what additional heating is required to 
drive a wind at time $t_n$?
To answer this question, we heated the gas 
by increasing the Type Ia 
supernova rate in NGC 4472 above the reference 
value SNu$(t_n) = 0.06$ SNu and repeated the calculation 
with all other parameters (including $s$) unchanged.
The transition to a wind is abrupt.
As seen in column 3 of Figure 5, for SNu$(t_n) = 0.18$
the temperature and density profiles at $t_n$ are almost 
identical to the reference solution,
although $M_{cold} = 2.5 \times 10^{10}$ $M_{\odot}$, 
$M_{hot} = 2.4 \times 10^{10}$ $M_{\odot}$ 
and $L_{x,bol} = 1.0 \times 10^{42}$ erg s$^{-1}$ are all 
slightly lower due to outflows at early times.
At time $t_n$ the gas is flowing inward at all radii.
However, a further small increase to SNu$(t_n) = 0.25$
produces a strong global wind at $t_n$ with very low gas density 
at all radii and $L_{x,bol}$ drops to $1.9 \times 10^{41}$ 
erg s$^{-1}$.
No known galaxy has density, temperature and abundance 
profiles like those for the SNu$(t_n) = 0.25$ solution 
in Figure 5.
Outflows generally require finely-tuned heating.
More realistic 
galactic flows with additional circumgalactic gas require  
a much larger SNu$(t_n)$ to drive an 
outflow by $t_n$. 
Outflows may be common 
in low luminosity ellipticals, $L_b \lta L_{B,crit}$, 
and spiral bulges where the hot gas is difficult to observe. 
If the reference Type Ia rate SNu$(t_n) = 0.06$ SNu 
is applied to  
elliptical galaxies that are $\lta 0.3$ as luminous as 
NGC 4472, outflows at $t_n$ are easy to generate.

{\bf Past Supernova Rate:}
Next we describe several flows at $t_n$ for 
a variety of past supernova rates SNu$(t) \propto t^{-s}$, 
by varying the index $s$ from 0 to 2, 
keeping SNu$(t_n) = 0.06$ fixed.
In row 3, column 4 of Figure 5 we show the current ($t = t_n$)
hot gas iron abundance profiles
for flows with $s = 0$, 1.5 and 2.
The temperature and density profiles at $t_n$ 
change little over this range of $s$.
$M_{cold}$ decreases by about 4.5 as $s$ increases from 0 to 2 
because of supernova-driven outflows at early times. 
$L_{x,bol}(t_n)$ only decreases by 40 percent 
as $s$ varies from 0 to 2.
Recall that $s > 1.3$ ($s < 1.3$) 
is a necessary condition for supernova 
driven winds to occur at early (late) times. 

Ciotti et al. (1991) considered evolutionary models 
with $s = 1.5$ and SNu$(t_n) = 0.11 - 0.22$ SNu, 
so that outflows and 
winds occur at early times, thereby reducing $M_{cool}(t_n)$. 
They also assumed that the fraction of galactic mass in 
dark halos varies among elliptical galaxies with 
similar $L_B$.
As a result, the model ellipticals described by 
Ciotti et al. are at the present time 
in different phases of a transition 
from outflow (low $L_{x,bol}/L_B$) 
to cooling inflow (high $L_{x,bol}/L_B$) and  
they interpret this as an explanation for the large scatter 
in the $L_x,L_B$ plot. 
However, for the same range of supernova parameters, 
the variations in $L_x/L_B$ would be greatly reduced 
if circumgalactic gas had been included.
Furthermore, the iron abundance at $t_n$ 
is strongly linked to the past 
supernova rate 
(Loewenstein \& Mathews 1991; Brighenti \& Mathews 1999b).
Type Ia rates required to drive winds at early times 
deposit too much iron in the hot gas by time $t_n$. 
The iron abundance for our $s = 1.5$ galaxy in 
Figure 5 is $\sim 3.5$ solar throughout the 
hot gas. 
This high abundance cannot be satisfactorily reduced 
by mixing with inflowing circumgalactic gas
(column 1, Figure 5).
We conclude that 
the transient evolution from supernova-driven 
winds to inflows is unlikely 
to explain the large scatter in the $L_x,L_B$ plot 
for massive ellipticals ($L_B \gta 3 \times 10^{10}$ 
$L_{B,\odot}$).

{\bf Additional Parameters:}
Finally, we note several additional parameters that have some  
influence on the solutions at time $t_n = 13$ Gyrs.
Varying the (uniform) stellar temperature $T_*$ from 
$6 \times 10^6$ K (the reference flow value) 
to $2 \times 10^6$ K or $18 \times 10^6$ K 
(but keeping the stellar mass fixed) 
leaves $n_e(r)$ and $T(r)$ essentially unchanged.
Matsushita (2001) argues
that $L_x$ and the hot gas temperature 
in E galaxies without circumgalactic gas 
can be explained by
kinematical heating of the gas by stellar mass loss. 
However, the kinematic temperature $T_*$ of 
mass lost from orbiting stars is essentially
the same as the hot gas temperature, 
i.e., both are determined by the same gravitational potential.
Altering the radiative cooling coefficient 
$\Lambda(T)$ by factors 3 or 1/3 has 
no appreciable effect on $T(r)$ but the gas density 
increases slightly with decreasing $\Lambda$.
If $\Lambda = 0$, as if some source of heating 
perfectly balances radiative cooling at every radius, 
then there is no radial flow and no gas cools at any radius.  
But in this strange solution the gas density becomes very large 
and the density gradient 
steepens ($\rho \propto \rho_*$) because a larger pressure gradient is 
required to support the denser, nearly static atmosphere.
Decreasing the mass of the NFW halo from 
the reference value $M_h = 4 \times 10^{13}$ $M_{\odot}$
to $M_{h} = 5 \times 10^{12}$ $M_{\odot}$ 
leaves $n_e(r)$ essentially unchanged 
but $T(r)$ is $\sim 20$ percent lower at $r = 10$ kpc.
For $M_h = 10^{12}$ $M_{\odot}$, however,
a mild galactic wind sets in at late times and 
the density and temperature are both significantly lower.
In this last solution $L_{x,bol}$ is $\sim 30$ times 
lower than the reference value and comparable to
the luminosity of discrete stellar X-ray sources 
in NGC 4472, $L_{x,*} \approx 3 \times 10^{40}$ erg s$^{-1}$
(Figure 1). 
It is also interesting to vary the stellar mass loss rate 
$\alpha_*$, simulating a situation in which not all of the 
gas ejected from stars goes into the hot phase.
When $\alpha_*(t_n)$ is reduced by 2, 
holding $\alpha_{sn}$ at its reference value,  
the total mass of cooled gas 
$M_{cold}$ and $L_{x}$ are both lowered by 2. 
The gas temperature within $\sim 40$ kpc 
is lower by $\sim 1.4$, but the total mass of hot 
gas $M_{hot} = 1.1 \times 10^{10}$ $M_{\odot}$ is almost 
unchanged.
If $\alpha_*(t_n)$ is lowered further, a wind develops 
by time $t_n$.

\section{THE COOLING PARADOX -- DOES THE GAS COOL?}

The cooling paradox refers to the remarkable discovery that 
the hot gas radiates but does not seem to cool. 
Many who have been 
closely involved with so-called ``cooling flows'' now 
doubt that they are cooling and some question whether they 
are flowing. 
Current attempts to resolve this paradox are generally taking 
two approaches: (1) the gas is cooling, but for some reason 
it evades detection and/or (2) the gas is being heated in 
some way so that very little gas cools. 
In this section we briefly review the evidence 
supporting the first approach, and heating is discussed in the
following section.

\subsection{Spectroscopic and Morphological Cooling Rates}

Most of the evidence that cooling 
is incomplete or absent comes from XMM RGS 
(Reflection Grating Spectrometer) spectra 
of cluster scale flows 
(Abell 1835: Peterson et al. 2001; 
Abell 1795: Tamura et al. 2001; 
Abell S1101: Kaastra et al. 2001).
In general the older data from ROSAT and ASCA could 
not easily distinguish between flows with a radially varying
single temperature (i.e. single phase)
and truly multiphase flows in which a range of 
temperatures is also present at every radius. 
For Abell 1835, a massive cluster with 
$T_{vir} \approx 9$ keV, Allen et al. (1996) estimated 
a large mass deposition rate,  
${\dot M} \approx 2000 $ $M_{\odot}$ yr$^{-1}$, 
using an X-ray image deprojection procedure based 
on a steady state cooling flow model.
This cooling rate has now been 
adjusted downward to ${\dot M} < 200$ $M_{\odot}$ yr$^{-1}$ 
as a result of the XMM RGS spectrum of Abell 1835 that shows no 
emission lines from ions at temperatures 
$T \lta T_{vir}/3$ (Peterson et al. 2001).
Similar drastic reductions were made to 
morphological ${\dot M}$ estimates for 
other clusters, although the RGS only views the 
central 1' of each cluster, a radius of 30''.
In these clusters, the gas temperature drops from 
$T_{vir}$ at large radii to about $\sim T_{vir}/3$ 
near the center, similar to Figure 2b, 
with little or no indication of multiphase (cooling) gas at 
lower temperatures, $kT \lta 1 - 2$ keV.

At present the XMM and Chandra 
data for galaxy/group flows are limited, but 
the spectroscopic ${\dot M}$ also seems to be 
significantly lower than the morphological ${\dot M}$, 
as in cluster flows.
XMM RGS observations of NGC 4636 by Xu et al. (2002) 
indicate that the gas temperature decreases from 
$8.1 \times 10^6$ K at 3 kpc to 
$6.4 \times 10^6$ K at the center, 
but there is little or no spectroscopic 
evidence for gas emitting at temperatures $\lta 6 \times 10^6$ K.
The OVII line at 0.574 keV that emits near $2.5 \times 10^6$ K is 
not detected.
Xu et al. determine
${\dot M} < 0.3$ $M_{\odot}$ yr$^{-1}$ within $r \sim 2.5$ kpc.
This is less than 
the total stellar mass loss rate expected in NGC 4636, 
$\alpha M_{*t} \approx 0.5$ $M_{\odot}$ yr$^{-1}$, but 
most of the gas in NGC 4636 is circumgalactic so the mass cooling 
rate expected in a traditional cooling flow model would be larger: 
${\dot M} \approx 1 - 2$ $M_{\odot}$ yr$^{-1}$   
at distance $d = 17$ Mpc (e.g. Bertin \& Toniazzo 1995). 
However, Bregman, Miller \& Irwin (2001) have observed
NGC 4636 with the Far Ultraviolet Spectroscopic Explorer (FUSE) 
and detected 
the OVI 1032,1038\AA~ doublet 
emitted from gas at $T \sim 3 \times 10^5$ K. 
It is generally assumed that this emission arises
from gas that is cooling, not from the stellar ejecta
as it is being heated to the hot gas phase.
The heating process, possibly involving shocks and 
rapid thermal mixing, is likely to be faster and more efficient 
than collisional excitation 
of emission lines during cooling 
(Fabian, Canizares, \& B\"ohringer 1994).
The FUSE observations indicate a cooling rate of only 
${\dot M} \approx 0.17 \pm 0.02$ $M_{\odot}$ yr$^{-1}$ 
for $d = 17$ Mpc.
Both XMM RGS and FUSE observations are consistent with 
a cooling rate that is $\sim 5$ times less than 
predicted from cooling flow models. 
However, FUSE has only a 30'' (2.5 kpc) square aperture 
containing $\sim 0.083$ of the total X-ray luminosity. 
If the cooling were spatially distributed in proportion
to the X-ray surface brightness, as reckoned by
Bregman et al., the total cooling rate in NGC 4636 would be
$\sim 0.17/0.083 \sim 2$ $M_{\odot}$ yr$^{-1}$, 
which is consistent with traditional cooling flow models.
The OVI line width observed with FUSE is 44 km s$^{-1}$. 
However, if the OVI emission shares the same 
kinematic widths $\sim 150$ km s$^{-1}$ as 
the H$\alpha$ + [NII] lines observed in the central
15'' of NGC 4636 by Caon et al. (2000), 
the additional flux in the broad wings 
might not have been detected by FUSE.

Spectroscopic X-ray and FUSE 
cooling rates for M87 and NGC 1399, 
respectively centered in the Virgo and Fornax 
clusters, are also much less than previously expected
morphological rates.
XMM-EPIC observations of M87 
indicate ${\dot M} \lta 0.5$ $M_{\odot}$
yr$^{-1}$ (Molendi \& Pizzolato 2001; Matsushita et al. 2002), 
much less than earlier predictions from X-ray images: 
$\sim 40$ $M_{\odot}$ yr$^{-1}$
(Peres et al.  1998)
and $\sim 10$ $M_{\odot}$ yr$^{-1}$
(Stewart et al. 1984; Fabian et al. 1984b).
In NGC 1399 Buote (2002) found that the XMM spectrum 
within $\sim 20$ kpc 
is best fit with a two temperature emission model; 
spectral models with 
a single temperature or a continuum of temperatures 
(as in a cooling flow) give less satisfactory fits.
Buote identifies the hotter phase (1.5 - 1.8 keV) 
with gas from the Fornax cluster 
and the colder phase (1.0 - 1.3 keV) 
with gas ejected from stars in NGC 1399, 
but there is no evidence that the cool phase cools 
to even lower temperatures. 
This is supported by FUSE observations 
of NGC 1399 by Brown et al. (2002) 
who find ${\dot M} < 0.2$ $M_{\odot}$ yr$^{-1}$
within a $3 \times 3$ kpc region at the center 
of NGC 1399 -- this is much less than 
the total cooling rate of $\sim 2$ $M_{\odot}$ yr$^{-1}$ 
inferred from the X-ray image using cooling flow models 
(Bertin \& Toniazzo 1995; Rangarajan et al. 1995). 
Perhaps the cooling is intermittent.

\subsection{Warm and Cold Gas at $T < 10^5$ K}

In view of the failure to detect significant cooling 
at X-ray temperatures, it is remarkable that faint, diffuse,
warm gas at $T \sim 10^4$ K (H$\alpha$ + [NII]) 
is almost universally observed near the centers of 
groups and clusters formerly identified as cooling flows. 
Although the mass of warm gas
($10^4 - 10^5$ $M_{\odot}$) is very much
less than the mass that would have cooled in the
cooling flow scenario, it could represent an important 
short-lived phase in the cooling process. 
The density of this warm gas, estimated 
from the [S II] 6716/6731 ratio, is $\sim 100$ cm$^{-3}$
(Heckman et al 1989; Donahue \& Voit 1997) 
compatible with pressure equilibrium with the hot gas 
within $\sim 1$ kpc of the galactic centers. 
The warm gas can be ionized by galactic starlight
from old, hot post-AGB stars
(Binette et al 1994) or 
by UV from locally cooling gas and thermal conduction 
(Donahue \& Voit 1991; 1997).
The total luminosity in H$\alpha$ + [NII] emission is
$\sim 1 - 10 \times 10^{39}$ erg s$^{-1}$. 
In E galaxies that are strong radio sources 
most of the line emission is concentrated within $\sim 1$ 
arcsecond of the center (Kleijn et al. 2002), 
but it is unclear if this is true in general.

Recent surveys of H$\alpha$ + [NII] images and 
kinematics in elliptical galaxies 
(Shields 1991; Buson et al. 1993; Goudfrooij et al. 1994; 
Macchetto et al. 1996; Caon, Macchetto \& Pastoriza 2001)
reveal that the images are rarely as smooth and regular 
as the stellar isophotes. 
The spatial irregularities 
in the H$\alpha$ + [NII] images appear to be compatible 
with the random H$\alpha$ + [NII] velocities 
$\sim \pm 150$ km s$^{-1}$ observed 
(Zeilinger et al. 1996; Caon et al. 2000). 
The warm gas clearly does not track the 
smooth rotation or dispersion kinematics 
of the stars but instead 
may approximately follow the motion of the hot gas.
If so, the hot gas would be subsonically turbulent 
with energy on spatial scales ($\sim 0.5$ kpc) 
too large to be produced by Type Ia supernovae.

A small number of elliptical galaxies have detectable 
neutral or molecular gas.
Surveys (Knapp, Turner \& Cunniffe 1985;
Huchtmeier 1994; Huchtmeier, Sage \& Henkel 1995) 
detect HI emission in $\lta 15$ percent of E galaxies, but 
in some cases this may include emission from nearby late type 
galaxies within the beam. 
HI disks are more 
common in low luminosity E galaxies that do not have 
observable X-ray emission from hot gas 
(Sadler, et al. 2000; Oosterloo et al. 2002). 
Some of the more spectacular examples, 
in which the HI gas probably results from a recent merger, 
have blue colors like spiral galaxies. 
Similarly, Wiklind, Combes \& Henkel (1995), 
Knapp \& Rupen (1996) and Young (2002) detect 
CO emission from a relatively small number of 
E galaxies selected because of their large 
far infrared fluxes. 
The molecular gas is often 
rotating in large ($1 - 10$ kpc) disks of mass $10^7 - 
10^9$ $M_{\odot}$. 
In some cases the high specific angular momentum 
of the molecular gas (or its sign) requires an external origin.
Using spectroscopic stellar ages, 
Georgakakis et al. (2001) have shown 
that the mass of neutral 
and molecular gas decreases with time 
(by star formation?) since the 
last merger event. 

\subsection{Dust in Elliptical Galaxies}

About $80$ percent of elliptical galaxies
have dust clouds, 
lanes or disks within $\sim 1$ kpc
of the center (van Dokkum \& Franx 1995;
Martel et al. 2000;
Colbert, Mulchaey \& Zabludoff 2001;
Tran et al. 2001; Kleijn et al. 2002) 
with associated gas masses $10^4 - 10^7$ $M_{\odot}$.
Some of the H$\alpha$-emitting gas is undoubtedly related 
to the photoionized boundaries of these dusty clouds 
(Goudfrooij et al. 1994; Ferrari et al. 1999).
One might expect the dusty cores having an irregular (non-disk)
appearance to be related to recent AGN outbursts,
but the evidence for this is not yet compelling
(e.g. Tomita et al. 2000; Krajnovic \& Jaffe 2002).
If AGN energy outbursts are common in E galaxies, 
it is remarkable that these rather fragile central dusty regions  
have survived.
IRAS luminosities, 
detected in $\sim 50$ percent of ellipticals, 
indicate that there is much more dust in E galaxies 
than that responsible for the optical extinction 
in the central clouds 
(Goudfrooij \& de Jong 1995; Bregman et al. 1998: 
Merluzzi 1998; Ferrari et al. 2002).
The ratio of 100 $\mu$m to optical B-band fluxes is 
noticeably higher for cD galaxies (Bregman, McNamara \& 
O'Connell 1990).
Knapp, Gunn and Wynn-Williams (1992) 
have shown that the mid-infrared 
$\sim 12$ $\mu$m radiation has an $r^{1/4}$ distribution 
like the galactic stars, consistent with circumstellar 
dust associated with stellar mass loss.
Recent midIR ISO spectra show the  
9.7 $\mu$m feature characteristic 
of circumstellar silicate grains expected in low mass AGB stars 
(Athey et al. 2002), this is proof that galactic stars 
are ejecting $\alpha$-elements.
According to Athey et al.,
the total midIR emission is consistent with typical 
stellar mass loss rates $\sim 1$ $M_{\odot}$ yr$^{-1}$ 
in luminous E galaxies, although ${\dot M}$ 
may not scale with $L_B$. 

\subsection{Origin of Warm Gas, Cold Gas and Dust}

The diffuse warm ($\sim 10^4$ K) 
gas that emits optical emission lines gas could have 
been recently ejected from stars, cooled from the hot phase, 
acquired in a recent galactic merger, or accreted from local gas 
clouds similar to the high velocity clouds 
that fall toward the Milky Way.
Interesting morphological similarities 
between the H$\alpha$ + [NII] and X-ray images 
have been reported in some E galaxies 
(Trinchieri \& di Serego Alighieri 1991;
Trinchieri, Noris, \& di Serego Alighieri 1997; 
Trinchieri \& Goudfrooij 2002), 
suggesting a generic relationship.
Optical line emission traces (part of) the perimeters 
of X-ray cavities in 3C317/Abell 2052 (Blanton et al. 2001b)
and Perseus/NGC 1275 (McNamara, O'Connell \& Sarazin 1996), 
but apparently does not fill in the cavities.
This may be significant. 
Since the cavity rims show little heating due to strong shocks, 
the pressure inside the cavity must be similar to that 
in the rims. 
If so, recent stellar ejecta or warm gas introduced 
by a merger 
should be just as luminous inside the 
cavities, but evidently this is not observed. 
Either the evaporation-heating time is shorter 
inside the cavities or   
the H$\alpha$-emitting gas may have cooled from the hot phase.

The total H$\alpha$ luminosity among luminous 
E galaxies, $L_{H\alpha} \approx 10^{39.5 \pm 1.0}$
ergs s$^{-1}$ (Goudfrooij et al. 1994), 
holds for E galaxies for which $L_x$ varies by $\sim 10^3$.
This uniformity 
and upper bound ($L_{H\alpha} \lta 10^{40.5}$ ergs s$^{-1}$) 
on the line emission may reflect the similarity of  
central hot gas pressure among massive E galaxies,
again suggesting (but not proving) an internal origin 
for the warm gas.
Mergers with gas-rich dwarfs might produce a wider 
spread in $L_{H\alpha}$.
Evidently all nearby X-ray bright E galaxies contain 
diffuse H$\alpha$ + [NII] emission.
It seems unlikely that mergers with gas-rich galaxies would 
be equally common for E galaxies in small groups and 
those in richer clusters.
The tendency for some of the diffuse H$\alpha$ + [NII] 
images to approximately 
follow the stellar isophotes (Caon et al. 2000)
also suggests an internal source for the warm gas, 
but the exceptions (e.g. NGC 5044) could arise from a merger or 
from recent AGN energy release.

The strange velocity patterns in the H$\alpha$ + [NII]
emission (e.g. Caon et al. 2001) 
may also constrain the origin of this warm gas.
The total stellar mass loss rate and hot gas 
cooling rate in classical ``cooling flows'' are both
$\sim 1$ $M_{\odot}$ yr$^{-1}$ in large E galaxies.
If this warm gas has a stellar origin,
its lifetime must be short, $t_{warm} \lta 10^5$ yrs,
to keep $L_{H\alpha}$ within observed limits
(Mathews \& Brighenti 1999a).
The velocity distribution of the 
warm gas is very different from that of the stars. 
But the warm gas clouds are 
$T_{hot}/T_{warm} \sim 1000$ times denser than the hot gas,
so they cannot be be drag-accelerated to the
local hot gas velocity in time $t_{warm}$.
Therefore the 
H$\alpha$ + [NII] emitting gas cannot be swept to the 
rims of the X-ray cavities by the motion of the hot gas. 
The chaotic velocity structure of the warm gas could be 
understood 
either as the undissipated motions of a very recent merger, 
or (more likely ?) as turbulent motions in the hot gas 
from which it cooled.
The obvious difficulty with this 
second possibility is that XMM observations so far 
provide little or no evidence for cooling.

It is often claimed that most or all of the 
cold gas and dust in E galaxies 
is a result of mergers with gas-rich galaxies
(e.g. Sparks et al. 1989; Sparks 1992; de Jong et al. 1990;
Zeilinger et al. 1996; Caon et al. 2000; 
Trinchieri \& Goudfrooij 2002).
If the warm gas has a finite lifetime, 
an ongoing supply of new gas is required.
However, there are relatively few reported cases 
of gas-rich dwarf galaxies currently merging with giant E
galaxies although 
the optical line emission would make such merging galaxies
easy to find. 
Moreover, most of the gas in merging galaxies could 
be lost by ram pressure stripping during the $10^8 - 10^9$ yrs 
as they orbit toward the center of the giant E galaxy. 
Undoubtedly, however, mergers do occur. 
A sure signature for mergers are the counter-rotating warm gas  
clouds found in some E galaxies by Zeilinger et al. (1996) 
and Caon et al. (2000). 
Further detailed 
observations of warm gas distributions and kinematics 
in normal E galaxies are needed to better constrain its origin. 

\subsection{Hiding the Cooling Gas}

The spectral contributions of cooling gas can be reduced if 
X-ray emission from the cooling gas is absorbed
or if the cooling is more rapid than normal radiative cooling
(Peterson et al. 2001; Fabian et al. 2001).

Intrinsic X-ray absorption has been invoked to interpret 
ROSAT and ASCA observations of cluster cooling flows 
(e.g. Allen 2000; Allen \& Fabian 1997; 
Allen et al. 2001). 
Column densities $N$ of a few $10^{21}$ cm$^2$ are sufficient 
to absorb 1 keV X-rays provided the absorbing gas is 
reasonably cold, $T \lta 10^6$ K. 
Because X-ray absorption varies rapidly with energy,
$\sigma(E) \propto E^{-3}$, it can reduce 
the low temperature contributions to 
the $\sim 1$ keV 
FeL iron line complex, simulating the absence of cooler gas 
(B\"ohringer et al. 2002). 
At ASCA resolution spectral models with conventional cooling flows 
are possible if the low temperature gas 
is largely hidden by intrinsic X-ray absorption. 
Buote (2000b,d; 2001) and Allen et al. (2001) 
reported evidence for an absorbing 
oxygen edge (at $\sim 0.5$ keV) and intrinsic absorbing columns 
$N \sim 10^{21}$ cm$^2$ within $\sim 5 - 10$ kpc
in several bright E galaxies such as NGC 5044 and NGC 1399.
But the mass of absorbing gas required to occult the central 
regions of these galactic flows is large,  
$M_{abs} \approx (4/3) \pi r^3 n m_p \sim 3 \times 10^7 r_{kpc}^2 N_{21}$
$M_{\odot}$ with unit filling factor.
It is unclear how this (necessarily colder) 
gas would be supported in the galactic gravity field and how 
it would escape detection at other frequencies 
(Voit \& Donahue 1995). 
Absorbing gas with $N \gta 10^{20}$ cm$^2$ would produce observable 
reddening of the background stars and large far-infrared luminosities 
unless the gas is dust-free. 
Resonant X-ray lines emitted from ions expected at 
low-temperatures can in principle be scattered and then absorbed 
by the continuous X-ray opacity (Gil'fanov et al. 1987), 
but the line-center optical depths are small  
and not all of the missing 
lines in XMM spectra are resonance transitions (Peterson et al. 2001).
XMM and Chandra spectra of the non-thermal nucleus and jet of M87 
and the nucleus of NGC 1275 fail 
to show intrinsic X-ray absorption at the level required 
to mask cooling in the thermal gas (B\"ohringer et al. 2001). 
Perhaps the ideal X-ray 
absorption model would be one in which the 
cooling regions are themselves optically thick at $\lta 1$ keV, 
but our attempts to achieve this have been unsuccessful.
Overall, the prospects for intrinsic X-ray absorption are not 
particularly encouraging.

If the cooling rate at $kT \lta 1$ keV were somehow accelerated, 
the X-ray emission from cooling gas would be reduced.
Begelman \& Fabian (1990) and Fabian et al. (2001)
propose that hot gas ($T_{hot}$) may rapidly mix with cold gas 
($T_{cold}$) and thermalize 
to $T_{mix} \sim (T_{hot} T_{cold})^{1/2}$ with little 
emission from temperatures $T_{hot} > T > T_{mix}$. 
Assuming $T_{hot} \sim 10^7$ K and $T_{cold} \sim 10^4$ K, 
then $T_{mix} \sim 3 \times 10^5$.
One potential difficulty with this process is that the $\sim 10^4$ K 
gas in E galaxies has a filling factor of 
only $\sim 10^{-6}$ 
(e.g. Mathews 1990) so it is not clear that the hot gas
can find enough cold gas to mix with. 
(Neutral and molecular gas at $T < 10^4$ K are also in short supply.)
In any case, when the gas at $T_{mix} \sim 3 \times 10^5$ K 
cools further, it should radiate 
strongly in the OVI UV doublet which, at least for NGC 1399, 
is not observed. 
Finally, 
the soft X-ray emission missing from cooling 
regions may appear somewhere else in the spectrum
(Fabian et al. 2002b).

Fujita, Fukumoto \& Okoshi (1997), Fabian et al. (2001) 
and Morris \& Fabian (2003) describe cooling flows  
in which the metal (iron) abundance is very inhomogeneous. 
Regions of enhanced iron abundance, possibly enriched by 
individual Type Ia supernovae, cool rapidly, 
reducing the X-ray line 
emission from intermediate temperatures, as required by XMM.
The remaining gas of lower metallicity can also  
cool with less line emission. 
This is an attractive idea that could help solve the 
cooling paradox, but it needs to be tested with detailed 
flow models.
In the presence of strong abundance inhomogeneities, 
the observed abundance and inferred Type Ia supernova rate 
will need to be appropriately adjusted.
This type of inhomogeneity would differ from successful models of
Milky Way enrichment where it is usually assumed
that Type Ia enrichment products
are widely distributed in the interstellar gas,
enriching subsequent stellar generations. 

Incomplete mixing of stellar ejecta can also have beneficial 
results.
Mathews \& Brighenti have recently modeled 
the thermal evolution of 
dusty gas ejected from red giant (AGB) stars in E galaxies 
in which the gas 
is first rapidly heated to the ambient temperature,
$\sim 1$ keV, then cools by thermal electron 
collisions with the grains. 
The electron-grain cooling time (Dwek \& Arendt 1992)
is faster than normal plasma cooling time,
the local galactic dynamical time and
the grain sputtering time.
Some grains survive in the cooled gas and may account for 
the central clouds of dusty gas observed in E galaxies.
Rapid electron-grain cooling may also be consistent 
with the weakness of
X-ray lines at subvirial temperatures in NGC 4636. 

In view of the small mass of gas with $T \lta 10^5$ K
in E galaxies, any gas that cools must rapidly 
convert to dense objects that are difficult to observe, 
like low mass stars. 
Star formation is aided by the high pressure and low 
temperature of the cold gas in E galaxy cores 
(Ferland, Fabian \& Johnstone 1994; 2002; 
Mathews \& Brighenti 1999a). 
In cluster cooling flows there is considerable 
evidence of ongoing star formation (Allen 1995; McNamara 1997;
Cardiel et al. 1998: Crawford et al. 1999;
Martel et al. 2002). 
For the much smaller cooling 
${\dot M}$ in E galaxies and their groups the evidence 
for star formation is less obvious such as the 
features in optical spectra from a ``frosting'' of young stars
(Mathews \& Brighenti 1999a; Trager et al. 2000). 

\section{COOLING FLOWS HEATED BY ACTIVE NUCLEI}

\subsection{Chandra Observations and Entropy Floors}

Instead of devising esoteric constructions 
to hide the cooling gas, why not simply 
heat the gas and avoid cooling altogether?
This is a good idea, but it is far from proven that 
it can work.
There is certainly enough energy in garden variety 
active galactic nuclei (AGN), $\sim \tau 10^{44}$ ergs s$^{-1}$, 
even for duty cycles $\tau$ of a few percent,
to balance the total energy radiated from galactic/group 
cooling flows, $L_x \sim 10^{41.5 \pm 1}$ ergs s$^{-1}$.  
The heating idea has received enormous 
additional support from 
high resolution Chandra X-ray images 
that reveal an amazing variety 
of structural disturbances near the flow centers 
where, according to current orthodoxy, 
supermassive black holes can be activated 
by inflowing gas to become AGNs.
So far most Chandra observations 
are of cluster-centered, E galaxies 
having strong radio sources, 
but similar hot gas irregularities
are also apparent in the centers of
more normal, group-centered
E galaxies and seem to be an almost universal phenomenon
[M84 (Finoguenov \& Jones 2001),
NGC 507 (Forman et al. 2001),
NGC 4636 (Jones et al. 2002, Loewenstein \& Mushotzky 2002),
NGC 5044 (Buote et al. 2003a). 

Of particular interest are the X-ray cavities, 
regions of much lower thermal emission that
are often coincident with radio lobes. 
In Hydra A/3C295
(McNamara et al. 2000; David et al. 2001;
Allen et al 2001) the non-thermal energy 
density in the radio lobes 
has apparently displaced the thermal gas.
In other cases, the radio-X-ray connection is less
clear: Perseus/NGC 1275
(B\"ohringer et al. 1993; Fabian et al 2000; Churazov et al. 2000),
A2052 (Blanton et al. 2001b) or A4059 (Heinz et al. 2002).
For example, 
only one of the two high-contrast central X-ray cavities in 
Perseus/NGC 1275 
is coincident with a maximum in the double-lobed radio emission.
The other radio lobe has apparently not displaced
the local thermal gas, although relativistic electrons
appear to be flowing from this radio lobe into 
an adjacent X-ray hole.
Perseus/NGC 1275 is a particularly interesting case because
it contains multitude of X-ray holes
located at random azimuthal orientations and distances
from the center.
Most of the ``ghost cavities'' at larger distances from the
galactic center no longer produce ($\sim$ GHz)
radio emission; either the relativistic electrons have
lost their energy or the holes have a different origin.
Whether the cavities are dominated by thermal or non-thermal plasma,
they must be buoyant on timescales $t_{buoy} \sim 10^7 - 10^8$ yrs
(e.g. Churazov et al. 2001). 
In any case, 
X-ray cavitation and other irregularities in the 
central hot gas suggest energy sources capable 
of heating the cooling flows.

Another manifestation of central heating may be the 
high entropy typically observed in 
group scale flows. 
In a perfect, starless, metal-free,  
$\Lambda$CDM, hierarchical universe
filled with adiabatic gas and self-similar dark halos,
the bolometric
X-ray bremsstrah\-lung luminosities of galaxy groups and
clusters would scale in a self-similar
fashion with gas temperature,
$L_x \propto T^2$ (Kaiser 1986; 1991; Evrard \& Henry 1991).
However, in our particular universe this relation is somewhat
steeper, $L_x \propto T^3$ (e.g. Arnaud \& Evrard 1999), 
and may become very steep
(at least $L_x \propto T^4$) for groups having
$kT \lta 1$ keV (Helsdon \& Ponman 2000a,b).
Evidently the gas in E-dominant groups has somehow acquired 
an additional ``entropy'' 
$S \equiv T/n_e^{2/3} \sim 100$ keV cm$^2$ 
in excess of that received in the cosmic accretion 
shock near the virial radius.  
This excess has been referred to as the ``entropy floor''
(Ponman et al. 1999; Lloyd-Davies et al. 2000). 
More recent studies 
(Mushotzky et al. 2003; Ponman et al. 2003) 
claim that the 
entropy at 0.1 of the virial radius varies as 
$S_{0.1} \propto T^{0.65}$ with no pause 
at the 100 kev cm$^2$ floor for groups and E galaxies;
$S_{0.1} \propto T$ would be expected in a perfect adiabatic 
universe.

The universality of 
entropy-enhancement in groups has led to the hypothesis that
the gas experienced some additional early heating
before (or as) it flowed into the group dark halos.
Early ``pre-heating'' by $\sim 1$ keV/particle could explain the
aberrant behavior of groups in both the $L_x-T$ and
$S-T$ plots.
Although this level of heating would only
be apparent in the shallow central potentials of galaxy groups,
many authors have adopted a stronger
assumption: that {\it all} baryonic gas, including gas
currently in both groups and clusters,
experienced some ``non-gravitational'' heating
(star formation, Pop III stars,
AGN, etc.)
prior to its entry into the dark halos 
(e.g. Tozzi \& Norman 2001; Borgani et al 2001, 2002). 
Heating at early cosmic times when the density is low 
is attractive because the entropy can be increased with 
a smaller energy expenditure per particle. 
The energy required exceeds that expected from 
Type II supernovae associated with the formation of 
the old stellar population in elliptical galaxies and 
spiral bulges, $\sim 0.2$ keV per particle, 
assuming a Salpeter IMF. 
Perhaps more attention should be paid to the possibility that 
much of the low entropy gas was removed by star formation and the 
remaining gas was heated 
by star formation (Loewenstein 2000; Brighenti \& Mathews 2001) 
and by a central AGN after the baryonic gas 
entered the dark halos.

\subsection{AGN Heating}

The absence of XMM spectral evidence for 
cooling at the expected level and obvious disturbances 
in Chandra images are the primary motivations for introducing 
some type of vigorous heating in the flow models. 
At the present time, however, both of these motivations 
are better documented for E galaxies in cluster environments than 
for similar galaxies in small groups that are more 
highly constrained.  
The gasdynamical consequences of heating have been 
investigated on two spatial scales: (1) attempts to arrest 
or retard radiative cooling throughout the hot gas and 
(2) studies of the evolution of small heated regions. 
We shall discuss these two approaches in turn.

In some models of heated ``cooling flows'' 
the physical heating mechanisms --  
shocks, cosmic rays, Compton heating, 
thermal conductivity etc. -- are 
explicitly invoked, but in others an 
{\it ad hoc} heating is simply imposed on the flow. 
Thermal conduction has been a popular 
heating option for many years
(e.g. Bertschinger \& Meiksin 1986; Meiksin 1988;
Bregman \& David 1988; Rosner \& Tucker 1989).
The uncertain influence of magnetic field geometry on the 
conductivity provides another adjustable parameter 
to fit the data.
If the thermal conduction term in Equation (5) 
is comparable to the radiative emission term,
$n_e^2 \Lambda \sim f 2 \kappa T/r^2$, 
the magnetic suppression factor must be 
$f \sim 56 (n_e r_{kpc})^2 T_7^{-7/2} 
\sim 0.2 r_{kpc}^{-0.36} T_7^{-7/2}$ 
where $n_e \approx 0.055 r_{kpc}^{-1.18}$ cm$^{-3}$ 
applies to NGC 4472 for $r_{kpc} \gta 0.3$ (Figure 2a).
This level of magnetic suppression is in excellent agreement with 
$f \sim 0.2$ determined by Narayan \& Medvedev (2001) 
for thermal conduction in a hot plasma with 
chaotic magnetic field fluctuations. 
Conduction could be very important 
in cluster-centered E galaxies such as M87 that are 
surrounded by much hotter gas;
these galaxies are of interest 
because they represent galactic scale flows with different 
outer boundary conditions. 
Regardless of the heating mechanism, it is essential 
that the density and temperature profiles 
in heated flows agree with observations. 
In particular, the heating must preserve the positive 
temperature gradients universally observed 
within several $R_e$ (Figure 2b).

\subsection{Cooling Flow Models with Heating}

Nonthermal
radiofrequency emission from cosmic ray electrons is observed
in the centers of 
$\sim 50$ percent of giant E galaxies 
(e.g. Krajnovic \& Jaffe 2002).
It is likely that proton cosmic rays are also
present or even dominant as in the Milky Way. 
Cosmic rays could heat the thermal gas 
either by Coulomb collisions 
(Rephaeli 1987; Rephaeli \& Silk 1995) 
or by dissipation of Alfven waves excited as 
cosmic rays stream along field lines
(e.g. Tucker \& Rosner 1983;
Rose et al. 1984;
Rosner \& Tucker 1989; 
Begelman \& Zweibel 1994).
Loewenstein, Zweibel \& Begelman (1991)
studied the propagation of cosmic
rays in a static hot gas galactic atmosphere having
an idealized globally radial magnetic field.
They suppose that outwardly streaming cosmic rays 
heat by wave dissipation at small galactic radii 
(also see B\"ohringer \& Morfill 1988)
whereas gas at larger radii is heated by thermal conduction.
Although their model is successful in producing
central regions with $dT/dr > 0$,
the central gas density gradient $dn/dr$
is flatter than observed
because the gas is both supported and heated by cosmic rays.
They conclude that this type of 
cosmic-ray heating cannot balance radiative cooling. 

Tabor \& Binney (1993) develop steady state flows
in which a convective core heated from the center 
is attached to an outer cooling inflow.
However, in the convective core the temperature gradient
is determined by the gravitational potential,
$(k /\mu m_p)(dT/dr) = -(2/5)(d \Phi /dr)$,
and must always be negative, contrary to observation.
The gas velocity in the convective core is problematical: 
if it is negative, all the gas
shed from evolving stars, $\sim 10^{10}$ $M_{\odot}$, 
would cool at the very center; 
if it is positive or zero, cooling would occur as a galactic 
drip (Mathews 1997) in compressed gas  
at the outer boundary of the convective core. 
Binney \& Tabor (1995) discuss time dependent 
hot gas flows in E galaxies without dark halos.
In addition to central AGN heating, 
a large number of Type Ia supernovae at
early times drive a galactic outflow
throughout much of the evolution 
and help to reduce the mass of cooled gas 
(as in Ciotti et al. 1991).
These supernovae are 
also likely to enrich the hot gas with iron
far in excess of observed abundances.
In the supernova heated flows 
$dT/dr$ is negative throughout the flow evolution,
unlike observed profiles, and 
the most successful models agree with 
observed gas density profiles only during a brief
$\sim 10^8$ year period preceding cooling catastrophes 
when the central gas density increases rapidly due to 
radiative losses. 
Binney \& Tabor also consider 
intermittent AGN heating within $\sim 1$ kpc triggered
by gas inflow into the center. 
As SNIa heating subsides and the flow evolves toward a
central cooling catastrophe, the AGN heating is activated
but the central gas density still continues to increase. 
As Binney \& Tabor suggest,
this may result in distributed cooling and
star formation in the central 1-2 kpc.
Such radiative cooling might not be possible in 
a strict interpretation of the XMM spectra. 

Tucker \& David (1997) study time dependent heated models
of cluster scale flows without cooling dropout or
mass supply from stars.
Gas near a central AGN is intermittently heated
by relativistic electrons when gas flows into the origin.
The temperature and entropy gradients
become negative and the density gradient is positive. 
Such flows are both convectively and Rayleigh-Taylor unstable.
When Tucker \& David include thermal conduction, 
the outward heat flux results in stable
solutions having cores of nearly flat density 
that are nearly isothermal. 
However, recent Chandra and XMM observations of 
the cluster-centered galaxies M87 in Virgo and 
NGC 4874 in Coma 
show that the gas temperature drops precipitously 
from 3 - 9 keV to $\sim 1$ keV between
50 and 10 kpc of the galactic centers
(B\"ohringer et al. 2001; Molendi \& Gastaldello 2001;
Arnaud et al. 2001; Vikhlinin et al. 2001).
These steep thermal gradients can be understood
with old-fashioned 
time-dependent cooling flow models with 
distributed cooling dropout 
(Brighenti \& Mathews 2002a), but only if 
thermal conductivity is suppressed ($f \lta 0.1$).

Ciotti \& Ostriker (1997; 2001) 
assume that gas in the cores of 
isolated elliptical galaxies is 
Compton heated by powerful isotropic gamma rays
($L_{bol} \sim 10^{46} - 10^{48}$ erg s$^{-1}$;
$T_C \approx \langle h \nu \rangle/4 \sim 10^9$ K)
whenever mass is accreted into an AGN at the center.
The explosive expansion of this ultrahot gas 
drives strong shocks out to 
$\sim 5$ kpc and beyond. 
The hard continuum is only activated about
$\sim 10^{-3}$ of the time, so only one in $\sim 1000$ of
known elliptical galaxies should be a gamma-dominant quasar
at any time.
For these models Ciotti \& Ostriker also assume
a high Type Ia supernova rate 
and no additional circumgalactic gas. 
Such hot gas atmospheres -- less deeply bound
and with small inertial masses -- 
can be significantly modified by AGN heating, 
reducing the mass of gas that cools near the galactic core.
However, $dT/dr$ is generally negative and
the gas density has a broad central core, 
although the density and temperature profiles
vary with time. 
The central core in the X-ray surface
brightness distribution varies
in size from 1 to 40 kpc which is much larger than
the $\sim 0.6$ kpc core in NGC 4649,
an elliptical having an $L_B$ similar to that
used in the Compton heated models.
It is unlikely, moreover, that gamma ray QSOs exist in 
all elliptical galaxies or that this radiation is isotropic.
AGN sources dominated by gamma rays, discovered 
with EGRET, are blazars or OVV (optically violent variables) 
type QSOs 
having flat radio spectra with superluminal VLBI activity,
suggesting that the hard radiation is highly collimated 
along the line of sight (e.g. von Montigny et al. 1995; 
Impey 1996; Hartman et al. 1999).
Finally, the Compton temperature $T_C$ 
of more typical, 
more isotropic AGNs and QSOs is less than $10^7$ K,
so cooling flow gas is more likely to be Compton cooled
than heated (Nulsen \& Fabian 2000).

Kritsuk, et al. (1998; 2001) 
describe 1D and 2D heated galactic flows having 
convective cores. 
The initial state for these flows is a static 
hot gas atmosphere in which the source terms 
in equations (3) and (5) dominate, 
i.e. gas ejected from stars is exactly consumed by 
distributed radiative dropout 
($-q \rho/t_{cool}$) and radiative losses 
are balanced by stellar and supernova heating
(Kritsuk, et al. 1998).
When additional central heating is supplied, the 
core of the flow becomes convective.
After $3 \times 10^8$ years the temperature 
gradient becomes negative in the convecting core 
a few kpc in size, unlike the profiles in Figure 2b. 
At this time an incipient cooling catastrophe 
seems to develop where the slowly outflowing 
convective core confronts the stationary gas beyond, 
similar to the thermal instabilities in 
galactic drips (Mathews 1997).
Brighenti \& Mathews (2002b) consider a wide variety of 
1D and 2D cooling flows that are heated within some 
radius by an unspecified process and triggered by 
gas flow into the origin.
Of the wide variety of heating scenarios considered,
none are in reasonable agreement
with observed hot gas temperature and density profiles.
Even for poor fits to the observations,
these models require finely tuned heating scenarios.
Idealized flows in which radiative cooling is perfectly
balanced by global
heating are grossly incompatible with observations. 
AGN feedback heating often results in spontaneous 
and spatially distributed cooling produced by 
non-linear compressions in turbulent regions.
Turbulence-induced cooling is 
associated with quasi-cyclic variations in the
hot gas density profile.
When cooling flows are partially supported 
by nonthermal pressure, similar cooling instabilities develop.
The global mass cooling rate is not altered by any form 
of heating considered -- in apparent violation of XMM spectra.

Ruszkowski \& Begelman (2002) describe a 
cluster flow model that is heated by conduction from
the outside and the entire flow is 
instantaneously heated 
when gas flows into the central AGN. 
The temperature gradient in these solutions is positive, 
and this much desired outcome is 
relatively insensitive to the scale 
and shape of the region heated by the central AGN. 
Successful Ruszkowski-Begelman flows require 
that the thermal conductivity suppression factor be in the  
range $0.1 \lta f \lta 0.5$. 
Fortunately this range includes the value $f \approx 0.2$ 
recently advocated by Narayan \& Medvedev (2001). 
Flows on galactic scales also benefit from the combined
effects of AGN heating and conduction, but the agreement
is generally less satisfactory than for cluster flows 
(Brighenti \& Mathews 2003).
Because of the lower temperature 
in galactic flows, $kT \sim 1$ keV, 
marginally acceptable, but not ideal, flow
solutions are possible only if the conductivity is close to  
its full Spitzer value, $f \approx 1$.
However, in these solutions the hot gas iron abundance is
several times solar throughout most of the flow. 
This is much higher than observed, even though the current Type Ia 
supernova rate used is low, SNu$(t_n) = 0.06$ SNu.
Therefore, to fully accept the beneficial combination of
heating plus conduction in galactic flows,
it is also necessary to hypothesize some means of reducing
the computed iron abundance. 
Some (but not all) of 
the iron produced by Type Ia may cool before entering
the hot gas, some selective cooling 
removes the excess iron, etc.

In summary, all current attempts to reduce the mass cooling rate 
${\dot M}$ in 
cooling flows by factors or 5 - 10 with various heating 
mechanisms are inadequate for 
one reason or another, particularly for flows in 
E-dominant group gas. 
When computed for many Gyrs, 
the heating often has little effect on ${\dot M}$ 
and the global density and temperature profiles 
disagree with the observations.
Thermal conduction has a helpful role, but is 
less effective in galactic flows. 
Essentially static flows cannot be correct either because 
the hot gas iron enrichment by Type Ia supernovae 
inside the central E galaxy 
would be enormous after several Gyrs.
Abundances are a major constraint on galactic flows. 

\subsection{Models of X-ray Cavities}

Several recent theoretical studies of X-ray cavities 
have examined
the consequences of introducing heated gas
in some localized
region away from the center of the cooling flow.
If the energy density in the cavities is sufficiently
large, they will expand supersonically,
producing shocks that heat the adjacent cooling flow gas 
(Clarke, Harris, \& Carilli 1997;
Heinz, Reynolds, \& Begelman 1998;
Begelman \& Cioffi 1989;
Rizza et al. 2000;
Reynolds, Heinz, \& Begelman 2001;
Soker, White, David \& McNamara 2001).
While such violent heating may occur in some situations,
recent Chandra and XMM observations
indicate that the gas temperature in the 
rims around the cavities 
and radio holes is typically cooler, not hotter, than average 
(Fabian 2001; McNamara 2001).
These low-entropy rims cannot be understood
as local gas that was shocked and subsequently lost
entropy by radiation
(Nulsen et al 2002; Soker, Blanton \& Sarazin 2002).

The 2D hot buoyant bubbles computed by
Churazov et al. (2001) slowly 
float upward in the cooling flow atmosphere
(also: Saxton et al. 2001; Br\"uggen \& Kaiser 2001;
Br\"uggen et al. 2002).
These rising cavities 
are accompanied by a column of colder (low entropy)
gas that moves radially upward near the center of the bubble.
Unless this colder gas is subsequently heated, however, 
it should eventually fall back.
Quilis et al. (2001) studied the evolution of a nearly axisymmetric
3D bubble produced by heated gas at some finite radius, 
simulating jet heating. 
They noticed that the buoyant bubble is surrounded
by a shell of slightly colder gas which they attributed
to cooling expansion as low-entropy gas is pushed by the
bubble toward regions of decreasing ambient pressure
(Nulsen et al 2002; Soker, Blanton \& Sarazin 2002).
Heating by jets 
can also create buoyant regions with slightly cooler rims 
that float upwards approximately along the jet axis
(Reynolds, Heinz, \& Begelman 2001; 2002).
Brighenti \& Mathews (2002c) show that 
central density irregularities and large, randomly
oriented X-ray holes may be a natural result of
the evolution of a single
spherically heated region at the center of a massive
E galaxy.
The gaseous rims around the holes
are cool, as observed, provided the heating occurs 
in the region of low entropy 
near the center of the flow 
and this is illustrated with an analogous similarity solution. 

\section{ROTATION AND MAGNETIC FIELDS}

If the classical cooling inflow hypothesis is correct, 
both rotation and magnetic fields should be amplified near the 
center of the flow.
Mass and angular momentum conservation in the hot gas 
within slowly rotating elliptical galaxies 
ultimately results in 
extended, massive disks of rotationally supported cold
gas in the equatorial plane. 
As the hot interstellar gas approaches
the disks, its density and thermal X-ray emission increase, resulting
in X-ray images that are considerably flattened toward the equatorial
plane out to an effective radius or beyond 
(Kley \& Mathews 1995; Brighenti \& Mathews 1996; 1997b). 
Such X-ray disks have never been observed 
(e.g. Hanlan \& Bregman 2000). 
This problem can be avoided if the gas is flowing outward, but 
we have seen that this requires that the heating be fine tuned. 
Alternatively, cooling inflows that rotate can have 
nearly circular X-ray images 
(1) if the mass of hot gas is depleted by 
spatially extended (multiphase) mass dropout or, 
in view of XMM observations to the contrary, 
(2) if angular momentum is transported outward by subsonic turbulent 
diffusion (Brighenti \& Mathews 2000).
If the random velocities of the diffuse H$\alpha$ + N[II] lines,
$\sim 150$ km s$^{-1}$, are also present in the hot gas, 
they would be sufficient for this diffusion.
Type Ia supernova rates of interest can 
maintain the centrally observed iron abundance gradient
even in the presence of 
interstellar turbulence sufficient to circularize 
the X-ray image (Brighenti \& Mathews 2000, Br\"uggen 2002). 

Similarly, seed magnetic fields contained in the ejected envelopes 
of evolving stars or in relic radio sources can be 
amplified by the same interstellar turbulence. 
Ancient fields established by turbulent dynamos during the merging 
epoch of the E galaxy group could still be present in the hot gas.  
Further amplification occurs as these fields
are compressed by an inward flow of hot gas
(Soker \& Sarazin 1990; Moss \& Shukurov 1996; 
Christodoulou \& Sarazin 1996; Mathews \& Brighenti 1997; 
Godon, Soker \& White 1998).
Field stresses of $\sim 50\mu$G, as observed in M87
(e.g. Owen, Eilek \& Keel 1990), 
formed in this way (without a central AGN) 
can compete with the thermal pressure in the hot gas 
and influence the observed gas density gradient.  
The generic incompatibility of magnetic fields and spherical 
gravitating atmospheres may 
give rise to some of the central X-ray surface 
brightness irregularities observed. 
Although potentially very important, studies of magnetic fields 
in cooling flows have been limited by the lack of observations 
in normal (weak radio) E galaxies and by the 
theoretical and numerical complexities of field reconnection.

\section{ABUNDANCES IN THE HOT GAS}

Abundance determinations in X-ray astronomy are confused 
by the ever-changing abundance of the sun. 
In much of the X-ray 
literature the ``photospheric'' iron abundance 
from Anders \& Grevesse (1989) 
is explicitly or implicitly assumed whereas the currently preferred  
``$\sim$meteoritic'' value $3.2 \times 10^{-5}$ (by number) 
is  lower by a factor 0.69
(McWilliam 1997; Grevesse \& Sauval 1999). 
Likewise, the solar abundance of O, S and Ar 
have recently 
changed by factors of 0.79, 1.32 and 0.69 respectively 
(Grevesse \& Sauval 1998) 
and this is offered as a new option in XPEC.
To facilitate comparisons between any two X-ray observations, 
it is essential that authors prominently 
list the solar abundances assumed, but this has not 
always been done. 

Evidently, at early times 
the hot gas associated with E galaxies
was enriched in metals by Type II
supernovae (SNII) that accompanied the formation of the
dominant old stellar population. 
At later times further enrichment has been provided 
by Type Ia supernovae (SNIa) and material expelled from 
red giant stars. 
For a proper cosmic census of supernova enrichment there 
is no substitute for rich clusters ($\langle kT \rangle \gta 3$ 
keV) that are massive enough to retain metal-enriched,
SNII-driven galactic winds (e.g. Renzini 1997; 1999). 
In group-centered elliptical galaxies 
most of the iron produced by SNII 
has been dispersed by a galactic wind 
into distant group gas or  
nearby intergalactic regions where the gas density is 
so low that it cannot be observed even with XMM 
(Brighenti \& Mathews 1999a). 
Groups are not closed boxes. 
Chandra observations by Martin, Kobulnicky \& Heckman (2002) 
of the dwarf starburst wind in NGC 1569 
verify that these winds transport almost all of the 
SNII enrichment into the local intergalactic medium 
(Vader 1986). 
Early epoch winds from both dwarf and giant galaxies 
were undoubtedly even stronger than 
those in local starburst galaxies. 
Ram-pressure stripping may also be relevant 
to hot gas enrichment (e.g. Toniazzo \& Schindler 2001).
Finally, ongoing metal-enriched galactic outflows 
energized by SNIa are energetically likely 
in low luminosity ellipticals, $L_B \lta L_{B,crit}$.

Yields from Type II supernovae $y_{II}$ 
(expressed in $M_{\odot})$ vary with initial main
sequence stellar mass (e.g. Hamuy 2003) and 
must be averaged over the initial mass function. 
While yields for elements that 
can be identified in X-ray spectra are still quite 
uncertain (e.g. Gibson, Loewenstein \& Mushotzky 1997;
Dupke \& White 2000; 
Gastaldello \& Molendi 2002), 
there are three broad categories: 
the ratio $y_{Ia}/\langle y_{II} \rangle$ is about 
10 for iron, $\sim 1$ for silicon and sulfur
(as well as argon, calcium and nickel), 
and $\sim 0.1 - 0.2$ for oxygen and magnesium.
Note that $y_{Ia}/\langle y_{II} \rangle$ is far
from uniform among the $\alpha$ elements
O, Mg, Ar, Ca, S, and Si.
Ideally, it should be possible to account for any 
observed abundance with a linear combination of 
SNII and SNIa yields, but current observations 
may be too uncertain to permit this. 
Abundances differ according to the quality of the observation,
the relative isolation of the K$\alpha$ or other 
prominent lines for a given element, 
the assumptions made about the 
distribution of gas temperature at every radius 
in the flow (1T, 2T, etc.),
and possibly also the plasma code (Gastaldello \& Molendi 2002).

In principle, abundances in the hot gas within the stellar images 
of E galaxies can provide critically important information 
about the direction and magnitude of the 
systematic radial flow of hot gas, $\rho u 4 \pi r^2$, 
which is also sensitive to distributed cooling. 
The known stellar density $\rho_*(r)$ 
and metallicity $z_*(r)$ define the 
source function for metal ejection. 
However, to extract this flow information one must 
combine uncertain observations with uncertain 
supernova yields and with uncertainties in the fraction 
of metal-enriched gas from SNIa and stellar winds 
that ultimately goes into the hot phase.

Nevertheless, most X-ray observers currently agree 
that the hot gas within and near group and 
cluster centered E galaxies has been 
enriched by SNIa 
(e.g. 
Matsumoto et al. 1996; 
Fukazawa et al. 1998; 
Dupke \& White 2000; 
Finoguenov \& Jones 2000;
Buote 2000a, 2002;
De Grandi \& Molendi 2001;
Buote et al. 2003b; 
Lewis, Stocke \& Boute 2002). 
The best observed E galaxy is 
M87 centered in the Virgo cluster 
(Matsumoto et al. 1996: 
Finoguenov et al. 2002; Gastaldello \& Molendi 2002), 
but the presence of a central AGN and X-ray jet 
in M87 may have disturbed the normal evolution of the hot gas.
Similar XMM observations of 
somewhat lower accuracy are available 
within a few 100 kpc of the   
central E galaxies in Abell 496 ($T \sim 2 - 5$ keV;
Tamura et al. 2001),
NGC 1399 (Buote 2002) and 
NGC 5044 ($T \sim 0.7 - 1.2$ keV; Buote et al. 2003b). 
No XMM abundances for NGC 4472 or 
other normal, group-centered E galaxies have been published 
as of the Fall of 2002. 
In all galaxies observed so far the hot gas iron abundance 
increases toward the central elliptical,
rising from $\sim 0.2 - 0.4$ solar at $\sim 100$ kpc 
to $\sim 1$ solar or more at the center, 
similar to the stellar abundances there.
[Older very subsolar central iron abundances determined 
with ASCA are now known to be wrong because 
the radial temperature variation was neglected 
(the ``Fe bias'': Trinchieri et al. 1994; 
Buote \& Fabian 1998; Buote 1999).] 
The scale of the iron enhanced region, $\sim 50 - 100$ kpc, 
is larger than the half-light radius $R_e$ of the 
central E (cD) galaxy.

Because oxygen and magnesium are formed in the outer layers of 
massive stars during normal stellar evolution,
their Type II supernova yields should be reliable
apart from uncertainties in the IMF and metallicity.
Consequently, one would expect O and Mg abundances 
to vary together.
Although XMM abundances of oxygen and magnesium are 
less accurate in the hot gas, they typically increase very little 
toward the core of the central elliptical and often become flat
or even decline near the very center
(Tamura et al. 2001; 
Buote et al. 2003b; 
Johnstone et al. 2002; 
Gastaldello \& Molendi 2002).
NGC 1399 and NGC 5044 data show that Mg traces 
Si and S more closely than O; only O seems flat. 
Either the O abundance in the mass-losing stars is surprisingly 
similar to that in the ambient hot gas, 
or the O expelled from O-rich stars is not all going into 
the hot gas.

The hot gas abundance $z_{i}$ 
(by mass in solar units) of element $i$ should vary as 
\begin{equation}
{d z_i \over d t} \equiv
{\partial z_i \over \partial t} 
+ u {\partial z_i \over \partial r}
= (z_{*,i} - z_i)\alpha_* {\rho_* \over \rho} 
+ {1.4 y_{Ia,i} \over z_{i,\odot} M_{sn}} 
\alpha_{sn} {\rho_* \over \rho}
\end{equation} 
where $y_{Ia,i}/M_{sn}$ is the fraction of 
mass in element $i$ ejected by Type Ia supernovae, 
$z_{i,\odot}$ is its solar abundance and 
1.4 is the ratio of total to hydrogen mass at 
solar abundance.
This equation is quite general, but it is interesting to 
apply it to subsonic galaxy/group inflows or outflows 
in which the gas density $\rho(r)$ remains essentially constant 
for several Gyrs.
We assume that all stellar and supernova 
ejecta goes into the hot gas.
In a Lagrangian frame moving slowly with 
the gas velocity $u$, the
abundance of element $i$ in the gas
increases from some initial value $z_{0,i}$
to an asymptotic limit
$$z_{eq,i}(r) =  z_{*,i}(r)
+ {1.4 y_{Ia,i} \over z_{i,\odot} M_{sn}}
{\alpha_{sn} \over \alpha_*}.$$
This equilibrium abundance is reached after 
$t_{eq} = (\rho/\rho_*)/\alpha_*
= 2.80 \times 10^8 r_{kpc}^{1.18}$ yrs,
using NGC 4472 densities from Figure 2a.

For iron enrichment ($y_{Ia,Fe} = 0.7$ $M_{\odot}$;
{$M_{sn} = 1.4$ $M_{\odot}$; SNu$(t_n) = 0.06$ SNu;
$z_{\odot,Fe} = 1.83 \times 10^{-3}$ by mass) we find
$z_{eq,Fe} = z_{*,Fe} + 2.35$ solar.
This value is similar to the gas abundance in
the isolated solution shown in the first column of
Figure 5 in which $z_{Fe}$ decreases from
$\sim 3$ solar at the center to
$\sim 2$ at large radii where $z_{*,Fe}$ becomes small.
It is significant that this flow solution
overestimates the iron abundance observed in NGC 4472, 
particularly for $r \gta 10$ kpc.
The correct iron abundance gradient in
$10 \lta r \lta 50$ kpc is, however, nicely fit
by the solution that includes circumgalactic gas
(dashed line, first column, Fig. 5).
This agreement is only possible 
if the low metallicity circumstellar (or cluster)
gas is flowing inward, diluting the entire flow, i.e.  
$z_{Fe}$ increases toward $z_{eq,Fe}$ 
as the gas flows in. 
Of course inflow 
is only possible if some (distributed) cooling occurs.
If the global radial flow is outward, 
the circumgalactic gas (with low $z_{Fe}$) 
would be pushed outward and 
the Fe abundance would approach $\sim z_{eq,Fe}$ within
$r \approx 3 t_{Gyr}^{0.85}$ kpc.
In this case the hot gas iron abundance $z_{Fe}(r)$ 
would be at least as large as that of 
the isolated flow solution (column 1, Fig. 5) 
which clearly exceeds the iron abundance observed in NGC 4472.
The observed iron abundance gradient is therefore 
a powerful argument for systematically  
inflowing hot gas, similar to the inflow required 
to explain the gas temperature profile ($dT/dr > 0$).

The abundance of oxygen in the hot gas, which comes mainly  
from stellar mass loss ($y_{Ia,O} \approx 0$), 
should approach the stellar abundance, 
$z_{eq,O} = z_{*,O}$ in time $t_{eq}$.
The O abundance in M87 rises gradually with $\rho_*$,
which is consistent with some stellar enrichment, but
it is curious that the O/Fe ratio is almost constant
with galactic radius (Gastaldello \& Molendi 2002). 
The small O abundance $z_O \sim 0.3$ 
observed in NGC 5044 and M87 
(Buote et al. 2003b; Gastaldello \& Molendi 2002) suggests that 
only a fraction of the gas lost from stars may go into 
the hot phase. 
Alternatively, the stellar O abundance in E galaxies may 
be subsolar, as suggested by observations of forbidden 
emission lines in E galaxy planetary 
nebulae (Jacoby \& Ciardullo 1999; Walsh et al. 1999), 
although nebular forbidden lines are known to give lower O 
abunances than the recombination lines. 
But O and Mg are expected to vary in a similar manner 
and stellar 
Mg is known to increase with $\rho_*$ (Trager et al. 2000). 
If the O abundance is systematically low in E galaxies, 
this may indicate the presence of hypernovae in which 
oxygen is burned into heavier elements (Loewenstein 2001).

One of the most remarkable 
peculiarities in the abundance profiles 
are the central minima observed for almost all 
elements in M87 (Gastaldello \& Molendi 2002), 
in NGC 4636 (Loewenstein \& Mushotzky), 
in AWM 7 (Johnstone et al. 2002) and elsewhere. 
For ellipticals that are resolved by ROSAT within 5 kpc, 
the observations compiled by
Buote (2000a,c) indicate that the iron abundance is often
flat (NGC 1399) or decreasing
(NGC 4472; NGC 5846; NGC 4636) within 5 kpc, 
assuming no intrinsic absorption. 
Only NGC 4649 shows an increase.
Subsolar central hot gas iron abundances, 
$\sim 0.8$ solar, for $r \lta 3$ kpc 
have been confirmed with XMM RGS observations 
(Xu et al. 2002; Sakelliou et al 2002).
It has been realized for some time that the central abundances 
could appear lower if the prominent X-ray resonance lines 
emitted there were optically thick 
and diffused outward before escaping 
(Gil'fanov et al. 1987, Tawara et al. 1997,
Shigeyama 1998; B\"{o}hringer et al. 2001;
Mathews et al. 2001; Sazonov et al. 2002). 
However, the XMM RGS spectra of Abell 496 
(Tamura et al. 2001) and M87 (Gastaldello \& Molendi 2002) 
show central reversals even for non-resonant lines.
In M87 the optical depth may be too low for line radiative transfer 
to be important, particularly if the hot gas is mildly 
turbulent 
(Mathews et al. 2001; 
Sakelliou et al. 2002).
Morris \& Fabian (2003) describe a transient enrichment 
process that produces a central minimum but this feature would 
probably disappear if the source function were constant 
with time. 
There is at present no fully satisfactory explanation of these 
strange abundance profiles, assuming they are real. 

We thank David Buote for reading an early draft of this 
review and for his helpful suggestions.
Studies of the evolution of hot gas in elliptical galaxies
at UC Santa Cruz are supported by
NASA grants NAG 5-8409 \& NAG 5-9956 and NSF grants
AST-9802994 \& AST-0098351 for which we are very grateful.
FB is supported in part by grant MURST-Cofin 00.

\listoffigures

\end{document}